\documentclass[english,11pt,oneside]{article}

\pdfoutput=1

\usepackage{amsmath}
\usepackage{amsfonts}
\usepackage{amssymb}
\usepackage{graphicx, rotating}
\usepackage{epstopdf}
\usepackage{epsfig}
\usepackage{latexsym}
\usepackage{multirow}
\usepackage{color}
\usepackage[dvipsnames]{xcolor}
\usepackage{slashed}
\usepackage[top=2.4 cm, bottom=2.4 cm, left=2.5 cm, right=2.5 cm]{geometry}
\usepackage{amstext}
\usepackage{array}  
\usepackage{hyperref}
\usepackage{scalerel}
\hypersetup{
colorlinks = true,
 linkcolor  = red, 
 linktocpage=true,
 citecolor={blue}
}
\usepackage{bbold}
\usepackage{physics}
\usepackage{mathtools}
\usepackage{feynmp-auto}

\usepackage[utf8]{inputenc}
\usepackage{bm}
\usepackage{leftidx}
\usepackage[numbers,sort&compress]{natbib}

\usepackage{makecell}

\usepackage{authblk}
\usepackage{colortbl}
\usepackage{xcolor}
\usepackage{enumitem}

\usepackage{float}

\usepackage[normalem]{ulem}
\usepackage{authblk}       
\usepackage{orcidlink}     
\usepackage{setspace}      

\newcommand{\be}{\begin{equation}}
\newcommand{\ee}{\end{equation}}
\newcommand{\ba}{\begin{array}}
\newcommand{\ea}{\end{array}}

\definecolor{LightGrey}{gray}{0.96}
\definecolor{Grey}{gray}{0.94}

\def\ang#1{{ \langle #1 \rangle}}

\begin{document}

\title{\bf 
Vector dark matter with non-abelian kinetic  mixing
}

\author{
Ana Luisa Foguel $^{1,}$\footnote{\href{mailto:afoguel@usp.br}{afoguel@usp.br}}       \orcidlink{0000-0002-4130-1200},
Renata Zukanovich Funchal $^{1,}$\footnote{\href{mailto:zukanov@if.usp.br}{zukanov@if.usp.br}}       \orcidlink{0000-0001-6749-0022} and
Michele Frigerio $^{2,}$$^{3,}$\footnote{\href{mailto:frigerio@lpthe.jussieu.fr}{frigerio@lpthe.jussieu.fr}}       
\orcidlink{0000-0002-8408-6117}
}

\affil{
$^{1}$ \emph{Departamento de F\'{\i}sica Matem\'atica, Instituto de F\'{\i}sica} \\
\emph{Universidade de S\~ao Paulo, C. P. 66.318, 05315-970 S\~ao Paulo, Brazil} \\
\vspace{5pt}
$^{2}$ \emph{Laboratoire Charles Coulomb (L2C), CNRS \& University of Montpellier, 34095 Montpellier, France} \\
\vspace{5pt}
$^3$ \emph{Laboratoire de Physique Théorique et Hautes Énergies (LPTHE),
CNRS \& Sorbonne Université, 4 Place Jussieu, Paris, France}}

\maketitle
\begin{abstract}
\noindent An appealing framework for dark matter is provided by light hidden sectors, below the electroweak scale, feebly coupled to the Standard Model via light mediators. We consider a minimal, predictive model where both the dark matter and the mediator are vector bosons, and have the same mass. 
The portal between the dark sector and the Standard Model is provided by a kinetic mixing between the dark gauge symmetry, $SU(2)_X$, and the hypercharge, $U(1)_Y$, induced by a dimension-six operator. 
The dark-matter candidates, $X^\pm$, are charged under a custodial symmetry and therefore stable, while the mediator is a massive dark photon, $Z_D$, mixing with the photon and the $Z$.
We show how the observed dark-matter abundance can be reproduced via freeze-out or freeze-in, through either the kinetic mixing or the dark gauge interaction. We also analyse dark 3-to-2 annihilations, that can become dominant in model variations with $Z_D$ heavier than $X^\pm$. We confront our relic-density predictions with current and projected experimental, astrophysical and cosmological bounds on the model parameter space, highlighting the correlation between the dark-photon and dark-matter phenomenologies.

\end{abstract}

\tableofcontents
\vspace{10pt}

\section{Introduction} \label{sec:intro}

The Standard Model (SM) of particle physics, while well established and remarkably successful, is missing a viable dark matter (DM) candidate. Its nature remains elusive
after nearly a century since the first astrophysical and cosmological evidences. For a long time, the DM search program focused on the hypothesis that the new particles would be heavy, with masses above the electroweak (EW) scale, requiring, in particular, high-energy colliders for their production. However, even after the advent of the LHC, no direct signs of heavy new physics beyond the Standard Model (BSM) have been observed. Furthermore, no signal of DM appeared in direct or indirect detection experiments, which have predominantly targeted such heavy mass range. This situation has motivated the possibility that the new dark physics could instead be light, though only feebly coupled to the SM, and therefore best tested in high-intensity precision experiments.

In order to build a model with a light, MeV-GeV scale DM candidate that is thermally produced in the early Universe, one must introduce a light mediator connecting the hidden sector to the SM. This dark portal is necessary since the available SM mediators, such as the weak gauge bosons, typically require heavier DM states with $m_{\rm DM} \gtrsim 2 \, {\rm GeV}$ (the Lee–Weinberg bound) in order to thermally reproduce the observed relic density. A renormalizable possibility to connect the SM to the dark sector through a single mediator is to introduce a new vector boson that kinetically mixes with the SM hypercharge gauge boson. The standard, minimal realisation of this vector portal requires a new, secluded $U(1)_X$ gauge symmetry (i.e.~all SM fields carry no $X$-charge), whose associated dark vector boson is traditionally known as the \emph{dark photon}~\cite{Holdom:1985ag}. This scenario, together with gauged $U(1)_Q$ models in which SM fields also carry a $Q$ charge (e.g.\ $Q = B-L$)~\cite{Ilten:2018crw,Bauer:2018onh,Foguel:2022ppx}  have been extensively studied in the literature (see e.g.~Ref.~\cite{Fabbrichesi:2020wbt} for a comprehensive review).

It is natural to ask which are the next-to-minimal models that can generate a kinetic mixing portal. An interesting possibility is to promote the dark gauge group from an abelian $U(1)_X$ to a non-abelian $SU(2)_X$, with a scalar field $\phi$ responsible for spontaneous symmetry breaking (SSB) and mass generation. In the literature, such an extension has mainly been explored in two distinct contexts. In~\cite{Hambye:2008bq,Baouche:2021wwa}, the triplet gauge bosons $X^a$ of $SU(2)_X$ serve as the DM candidate, with the correct relic abundance achieved through the Higgs portal only. In~\cite{Choi:2019zeb,Catena:2023use}, in order to introduce a renormalizable vector portal, the dark gauge group is taken to be $SU(2)_X \times U(1)_X$, with the abelian factor kinetically mixing with the hypercharge, as in the minimal models. A related framework with an $SU(2)$ dark gauge symmetry was analysed in~\cite{Chaffey:2019fec}, with focus on the DM self-interactions. Besides these constructions, non-abelian portals and spin-1 dark matter scenarios have been explored more broadly in the literature~\cite{Arkani-Hamed:2008hhe,Barello:2015bhq,Elor:2018xku,Delaunay:2020vdb}.

In this work we want to explore the kinetic mixing portal, without enlarging the dark gauge group $SU(2)_X$. To this end, we allow for non-renormalizable interactions, considering, in particular, the dimension-six operator $\mathcal{O} \sim (\phi^\dag \sigma^a \phi)\, X^a_{\mu\nu} B^{\mu\nu}$, with $\phi$ a dark scalar doublet and $X^a_{\mu\nu}$, $B_{\mu\nu}$ the gauge field strength tensors. After SSB of $SU(2)_X$, the usual custodial symmetry enforces  mass degeneracy among the three gauge bosons $X^a$. A residual global dark symmetry, $U(1)_D$, is preserved also by the dimension-six operator: the $U(1)_D$-charged gauge bosons $X^\pm$ remain stable and constitute spin-one DM candidates. Moreover, the dimension-six operator induces a kinetic mixing between the neutral vector boson $X^3$ and the SM hypercharge vector boson $B$, resulting, after canonical normalisation and mixing, in a new neutral mediator $Z_{D}$, that serves as the dark photon. Interestingly, we will see that the same operator also generates an effective DM-photon coupling, nonetheless the DM remains exactly neutral under electromagnetism.

The paper is organized as follows. In section \ref{sec:model}, we introduce in detail the minimal model for vector DM with kinetic mixing. 
Subsection~\ref{sec:mixings} is devoted to the identification of the relevant interactions for the DM dynamics, and the transformations to obtain the neutral-bosons mass eigenstates. 
In subsection~\ref{subsec:split}, we discuss variations of the minimal model with a non-negligible mass splitting between the DM candidates $X^\pm$ and the mediator $Z_D$.
In section~\ref{sec:relic}, we focus on the calculation of the DM relic abundance. Subsections~\ref{subsec:FO} and~\ref{subsec:FI} analyse the thermal freeze-out (FO) mechanism and the freeze-in (FI) mechanism, respectively. Subsection~\ref{subsec:FOsplit} considers the FO scenario in the presence of a mass splitting, to illustrate how the relative weight of different DM production channels changes. Section~\ref{sec:pheno} discusses the phenomenological consequences of the model, emphasizing how the kinetic-mixing portal provides a rich range of possibilities to produce the dark mediator $Z_D$, and test the FO and FI scenarios across different regions of parameters. We also comment on the interplay between the dark-photon searches and DM searches via direct and indirect detection, as well as DM self-interactions. We conclude with a summary of our main findings in section~\ref{sec:conclusion}.

\section{The vector dark matter model}\label{sec:model}

We consider a dark sector with a new $SU(2)_X$ gauge symmetry and the
triplet of gauge bosons $X^a_{\mu}$ ($a=1,2,3$) associated with it as well as a new scalar field $\phi$, in the group fundamental representation, which will endow them with a mass after SSB. The corresponding Lagrangian is given by
\be
{\cal L}_X = -\frac 14 X_{\mu\nu}^a X^{a\mu\nu} + (D_\mu \phi)^\dag (D^\mu\phi) -\lambda_X \left(\phi^\dag \phi -\frac{v_X^2}{2}\right)^2~, 
\ee
where
\be
X^a_{\mu\nu} \equiv \partial_\mu X^a_\nu - \partial_\nu X^a_\mu~ + g_X \, \varepsilon^{a b c} X^b_\mu  X^c_\nu~, \qquad
D_\mu \phi \equiv \qty(  \partial_\mu  -  i g_X
\frac{\sigma^a}{2} X^a_\mu) \phi~,
\ee
with $\varepsilon^{abc}$ the totally antisymmetric $SU(2)_X$  structure constants, $\sigma^a$ the Pauli matrices and $g_X$ the $SU(2)_X$ gauge coupling. As in the case of weak interactions, after spontaneous symmetry breaking at scale $v_X$, one is left with an accidental global symmetry, a custodial $SU(2)_D$. The three gauge bosons, $X^\pm_\mu \equiv (X^1_\mu \mp i X^2_\mu)/\sqrt{2}$ and $X^3_\mu$, form a custodial triplet that is degenerate in mass, while the scalar radial mode $h_X$ is a custodial singlet with mass $m_{h_X}^2 = 2\lambda_X v_X^2$.

The lowest dimensional operator that may couple such a dark sector to the SM is the Higgs portal, 
\be 
{\cal L}_{HP} = \lambda(\phi^\dag \phi)(\Phi^\dag \Phi)= \frac{\lambda}{4}(v_X+h_X)^2(v+h)^2~,
\ee
with $\Phi$ the SM Higgs doublet, $v\simeq 246$ GeV its vacuum expectation value (VEV), and $h$ the Higgs boson field.
The DM freeze-out through the Higgs portal has been extensively studied in prior literature~\cite{Hambye:2008bq,Baouche:2021wwa,Gross:2015cwa} and is subject to stringent constraints from various signatures, including Higgs invisible decays, triple Higgs coupling and di-Higgs production~\cite{Baouche:2021wwa}.

Here, we focus on the regime in which Higgs portal effects are subdominant. This can be realized either in the limit $m_{h_X}\gg m_X$ (i.e.~$\lambda_X\gg g_X^2$), since SM-DM interactions are mediated by $h_X$ exchange, or for a sufficiently small coupling $\lambda \ll 1$ (thermalisation with the SM bath is not efficient anymore for $\lambda \lesssim 10^{-7}$~\cite{Hambye:2008bq}). The smallness of 
$\lambda$ can be technically natural, e.g.~in extended UV models where $H$ and/or $\phi$ arise as pseudo-Nambu-Goldstone bosons, with $\lambda$ breaking the shift symmetry of each of the two fields~\cite{Frigerio:2012uc}.

Additional portals between the SM and the dark sector are provided by operators of dimension~$\ge~6$. In particular, a kinetic mixing between $SU(2)_X$  and the hypercharge symmetry $U(1)_Y$ may be induced, 
\be
{\cal L}_{KM} = \frac{1}{\Lambda^2} (\phi^\dag \sigma^a \phi) X^a_{\mu\nu} B^{\mu\nu} \supset -\frac{\epsilon}{2c_w} X^3_{\mu\nu} B^{\mu\nu}~,\qquad \epsilon\equiv \frac{{c_w}v_X^2}{\Lambda^2}~,
\label{KM}
\ee
where $\epsilon$ is the kinetic mixing parameter and $c_w\equiv \cos\theta_w$ is the cosine of the weak mixing angle.
One can check that this dimension-six operator breaks the global custodial $SU(2)_D$ symmetry but still preserves the subgroup $U(1)_D$ associated with the third generator. The states $X^{\pm}_\mu$ carry unit charge, while $X^3_\mu$ and $h_X$ are neutral. As a consequence, $X^\pm_\mu$ is a vector dark-matter candidate, whose stability is protected by $U(1)_D$, while $X^3_\mu$ can mix with the SM neutral vector bosons. Such higher-dimension operator can be generated, for instance, in a UV-complete scenario with heavy fields charged under both $SU(2)_X$ and $U(1)_Y$, see e.g.~\cite{Zhou:2022pom}.

Some comments are in order. First, the requirement of $U(1)_D$ invariance must be imposed on the UV-completion of the model, since, in general, other dimension-six operators can break $U(1)_D$ and destabilize dark matter. An example is the operator $(\tilde\phi^\dag\sigma^a \phi) X^a_{\mu\nu} B^{\mu\nu}$, with $\tilde\phi\equiv i\sigma_2\phi^*$. Second, in the SM the role of $U(1)_D$ is played by $U(1)_{\rm em}$ (the electric charge), but here it is not gauged. Third, within the SM Effective Field Theory (SMEFT), the analogue of eq.~\eqref{KM} is the dimension-six operator $(\Phi^\dag \sigma^a \Phi) W^a_{\mu\nu} B^{\mu\nu}$, which induces a contribution to the  oblique parameter $S$, that is constrained by EW precision tests. Accordingly, a similar constraint from the $Z^0$ precision measurements will hold on $\epsilon$~\cite{Loizos:2023xbj}. Finally, analogous operators where $B^{\mu\nu}$ is replaced by the dual field strength $\tilde B^{\mu\nu}$ can be neglected, as long as CP symmetry holds 
to good approximation.

\subsection{Masses, mixings and interactions of the new vector bosons}\label{sec:mixings}

Let us consider the three gauge bosons $X^3_\mu$, $B_\mu$ and $W^3_\mu$, associated to the generators $T^3_D$, $Y$ and $T^3$ respectively. They are neutral with respect to both conserved charges, $U(1)_D$ and $U(1)_{\rm em}$. 
In order to get the canonically normalized neutral mass eigenstates, one has to take into account the kinetic mixing term between $X^3$ and $B$, defined by eq.~\eqref{KM}, and the mass terms
\be
{\cal L}_m = \frac 12 m_X^2 (X^3)^2 + \frac 12 \hat m^2_{Z}
(c_w W^3-s_w B)^2~,\quad m^2_X=\frac{g_X^2v_X^2}{4}~,\quad \hat m^2_{Z}\equiv\frac{g^2v^2}{4\, c_w^2}~,
\label{Lm}\ee
where $g$ is the weak gauge coupling, and $c_w$ ($s_w$) the cosine (sine) of the weak mixing angle. The physical mass eigenstates are obtained through the following diagonalisation
\be 
\begin{pmatrix} Z_D \\ A \\ Z \end{pmatrix} =
\begin{pmatrix}
c_\alpha & 0 & -s_\alpha \\
0 & 1 & 0 \\
s_\alpha & 0 & c_\alpha \\
\end{pmatrix}
\begin{pmatrix}
 1 & 0 & 0 \\
 0 & c_w & s_w \\
 0 & -s_w & c_w \\
\end{pmatrix} 
\begin{pmatrix}
\sqrt{1- \frac{\epsilon^2}{c_w^2}} & 0 & 0 \\
 \frac{\epsilon}{c_w}  & 1 & 0 \\
 0 & 0 & 1 \\
\end{pmatrix} 
\begin{pmatrix} X^3 \\ B \\ W^3 \end{pmatrix}~,
\label{AZ}
\ee
where we have, from right to left, a non-orthogonal field redefinition to canonically normalize the kinetic terms, an orthogonal rotation to identify the massless photon field $A$, and a second orthogonal rotation, parameterised by the angle $\alpha$, to diagonalize the mass matrix of the two remaining states, thus defining two mass eigenstates $Z_D$ and $Z$.

The model has two free parameters relevant to the vector-boson masses and mixing: the kinetic mixing $\epsilon$ and the dark matter mass $m_X$. The third free parameter, the dark gauge coupling $g_X$, is relevant for the interactions among the mass eigenstates and will be discussed later. The SM parameters $\theta_w$ and $m_Z$ are fixed to their precisely measured experimental values. In the limit $\epsilon\ll 1$, as required by experimental constraints, one finds
\be 
\tan 2\alpha = \frac{2t_w}{1-m_X^2/m_Z^2} \epsilon +{\cal O}(\epsilon^3) ~,
\qquad 
\qquad
m_{Z_D}^2 = \left[1+\frac{\epsilon^2}{c_w^2} - \frac{t_w^2\epsilon^2}{1-m^2_X/m^2_Z}+{\cal O}(\epsilon^4)\right] m_X^2 ~,
\ee
where $t_w\equiv\tan\theta_w$. In what follows, we will mostly focus on the regime $m_X^2 \ll m_Z^2$. Note that in this minimal model the masses of $X$ and $Z_D$ are always quasi-degenerate, differing only by terms of order $\epsilon$. 
In appendix~\ref{details} we provide the exact relations (before $\epsilon$ expansion) connecting $\epsilon$ and $m_X$ to $m_{Z_D}^2$ and $\alpha$, and we also discuss the special case when $1 - m_X^2/m_Z^2 = {\cal O}(\epsilon)$.

Let us now describe the vector boson interactions relevant to the purposes of our paper. All couplings linear in the physical fields $A$, $Z_D$ and $Z$ can be obtained from the neutral currents,
\be
{\cal L}_{NC} = g_X J^\mu_{X^3} X_\mu^3 + g' J^\mu_B B_\mu + g J^\mu_{W^3} W^3_\mu~,
\ee
by applying the field redefinition of eq.~\eqref{AZ}. The dark current only contains the DM field,
\be 
J^\mu_{X^3}= i(2X^{-\nu}\partial_\nu X^{+\mu}+X^{+\mu} \partial_\nu X^{-\nu}
+X^{+\nu} \partial^\mu X^-_\nu)+h.c.~,
\label{cubic}\ee 
while the $B_\mu$ current contains the SM hypercharge current, $J_B^{SM,\mu}$, which includes all SM fermions and the Higgs, together with an additional DM contribution induced by the operator in eq.~\eqref{KM},
\be\label{eq:Aportal}
J_B^\mu = J_B^{SM,\mu} + \frac{g_X}{g'} \, J^{X,\mu}_{B} \,, \quad \quad \quad J^{X,\mu}_{B}=i \frac{\epsilon }{c_w}
\partial_\nu (X^{-\nu}X^{+\mu}-X^{+\nu}X^{-\mu})~.
\ee
Finally, $J^\mu_{W^3}= J^{SM,\mu}_{W^3}$ is the usual SM weak neutral current, including all SM weak doublet fermions, the Higgs, and the charged gauge bosons $W^\pm_\mu$ in analogy with eq.~\eqref{cubic}.

Interestingly, the current $J^{X,\mu}_{B}$ induces a DM coupling to the photon. At first sight, one might think that the DM acquires a millicharge of order 
$(\epsilon \, g_X)$. A closer inspection of  $J^{X,\mu}_{B}$ shows that this is not the case, as one can check that $X^\pm$ is {\it not} subject to a long-range  Coulomb potential in the non-relativistic limit.
Indeed, spin-one charged particles, such as the SM $W^\pm$, couple to the photon with a different Lorentz structure, analogous to the one in eq.~\eqref{cubic}. The DM-photon coupling induced by the operator in eq.~\eqref{KM} rather corresponds to a (tiny) magnetic dipole interaction (note that $W^\pm$ in the SM
has both a charge and a magnetic dipole).
Consequently, such photon portal is {\it not} subject to the constraints on millicharged DM candidates. The DM phenomenology of our model will be discussed in section~\ref{sec:pheno}.

In addition to the currents coupled to a single, neutral gauge boson, there are several interaction terms bilinear in $X^3$, $B$, and $W^3$. The only one relevant for our analysis is the dark sector quartic coupling involving two $X^3$ fields, given by
\be 
{\cal L}_{X} \supset g_X^2\, (X^3_\mu X^3_\nu X^{+\mu} X^{-\nu}
-X^3_\mu X^{3\mu} X^{+\nu} X^-_\nu)~.
\ee

\subsection{Model variations with a heavier mediator}\label{subsec:split}

The model described above is both minimal and predictive. In particular, the three dark gauge bosons, $X^\pm$ and $X^3$, are degenerate in mass. The kinetic mixing induces only a tiny mass splitting, of order $\epsilon$, between the dark matter candidates $X^\pm$ and the diagonalized physical mediator state $Z_D$ (see section \ref{sec:mixings} for details). One might wonder whether additional dimension-6 operators may lift this degeneracy, however, the corrections they generate to the gauge boson masses always scale as $\sim g_X^2 v_X^2/\Lambda^2$, and are therefore necessarily small within the regime of validity of the EFT. As a result, the near-degeneracy between the DM states $X^\pm$ and the mediator $Z_D$ is a robust feature of the minimal model, and a leitmotif of our main analysis.

Less minimal models may, nevertheless, allow for a significant mass splitting between $m_{X^3}$ and $m_{X^\pm}$. One possibility is to enlarge the gauge group to $SU(2)_X\times U(1)_X$, spontaneously broken in full analogy to the SM by a doublet $\phi$ carrying a non-zero $U(1)_X$ charge. In this scenario, there is an unbroken dark gauge symmetry, $U(1)_D$, which guaranties the stability of $X^\pm$, and implies the existence of a massless dark photon $A_D$. The heavy mediator mass reads $m_{X^3}^2= m_{X^\pm}^2(1+g_X'^2/g_X^2)$, where $g'_X$ is the $U(1)_X$ gauge coupling. Clearly, by raising $g'_X/g_X$ one can easily induce a large mass splitting. In principle, this setup allows for more kinetic mixing operators with respect to the minimal model, including a dimension-4 one between $U(1)_X$ and $U(1)_Y$. However, it is easy to conceive of a UV completion that induces only the operator in eq.~\eqref{KM}, e.g.~by introducing heavy states charged only under $SU(2)_X$ and $U(1)_Y$. The unavoidable difference, with respect to the minimal model, is the extra massless dark photon, with radically different phenomenology. 
Alternatively, one can realize the spontaneous breaking of the full $SU(2)_X\times U(1)_X$ symmetry, by introducing additional scalars, so that no massless dark photon remains. This scenario has been studied in some detail in the literature~\cite{Choi:2019zeb,Catena:2023use,Choi:2017zww}: the mass hierarchy among $X^{\pm}$, $X^3$, and the extra $Z'$, together with the usual kinetic mixing portal between $U(1)_X$ and $U(1)_Y$, can account for the observed DM abundance, with a rich associated phenomenology.

Another possibility to split $m_{X^3}$ and $m_{X^\pm}$ is to stick to the minimal $SU(2)_X$ gauge group while introducing non-fundamental representations of $\phi$ different than the fundamental. In the simplest case of a single real triplet, $\phi^a\sim \bf{3}$, its VEV breaks the symmetry to $U(1)$ leaving the neutral gauge boson massless. In contrast, in a model with two real triplets
$\phi^a_1,\phi^a_2\sim \bf{3}$, they can acquire VEVs $v_1$ and $v_2$ in orthogonal directions and thus break the $SU(2)_X$ symmetry completely, with gauge boson masses given by $g_X^2(v_1^2,v_2^2,v_1^2+v_2^2)$. Another interesting scenario occurs with one real fiveplet, $\phi^{ab}\sim \bf{5}$, parameterised as a $3\times 3$ symmetric and traceless matrix \cite{Brummer:2023znr,Brummer:2024ejc}. In this case, the VEV direction can vary in  field space as a function of an angular parameter, on which the gauge boson masses will depend in a non-trivial way~\cite{Brummer:2023znr,Brummer:2024ejc}. For a special value of the angle one obtains $m_{X^3}/m_{X^\pm} = 2$, a ratio which allows for resonant DM annihilation \cite{Nomura:2020zlm}. The latter reference also provides gauge boson masses for other scalar representations.

In the following, we first analyse in detail the DM relic abundance in the FO and FI scenarios for our minimal model with $R\equiv m_{X^3}/m_{X^\pm}=1$, and then we allow for an arbitrary mass ratio $R$ in order to study the impact of the mass splitting on the DM production channels.

\section{Relic density of the vector dark matter}\label{sec:relic}
The relic abundance of the dark matter candidates can be computed by solving the Boltzmann equation for the evolution of the DM number density $n_{X}$ as a function of temperature $T$. Here, we will assume only CP conserving processes in the dark sector, such that $n_{X^+} = n_{X^-}$ and we define $n_{X} \equiv n_{X^+} + n_{X^-}$.\footnote{The equality of $X^+$ and $X^-$ number densities is also guaranteed by the dark-charge symmetry $U(1)_D$, provided the Universe is initially neutral.}

The number-changing process contributing to the relic density abundance can be classified into three categories: i) $s$-channel $2 \to 2$ dark-to-SM processes via kinetic mixing, ii) $2 \to 2$ dark-to-dark processes driven by dark-sector self-interactions, and iii) $3 \to 2$ processes also driven by dark-sector self-interactions. We will sometimes refer to these regimes as the \textbf{kinetic mixing}, \textbf{dark annihilation}, and \textbf{SIMP} (for Strongly Interacting Massive Particles) regimes, respectively. 

\begin{figure}[tb!]
\begin{center}
\vspace{10pt}
\includegraphics[width=0.97\textwidth]{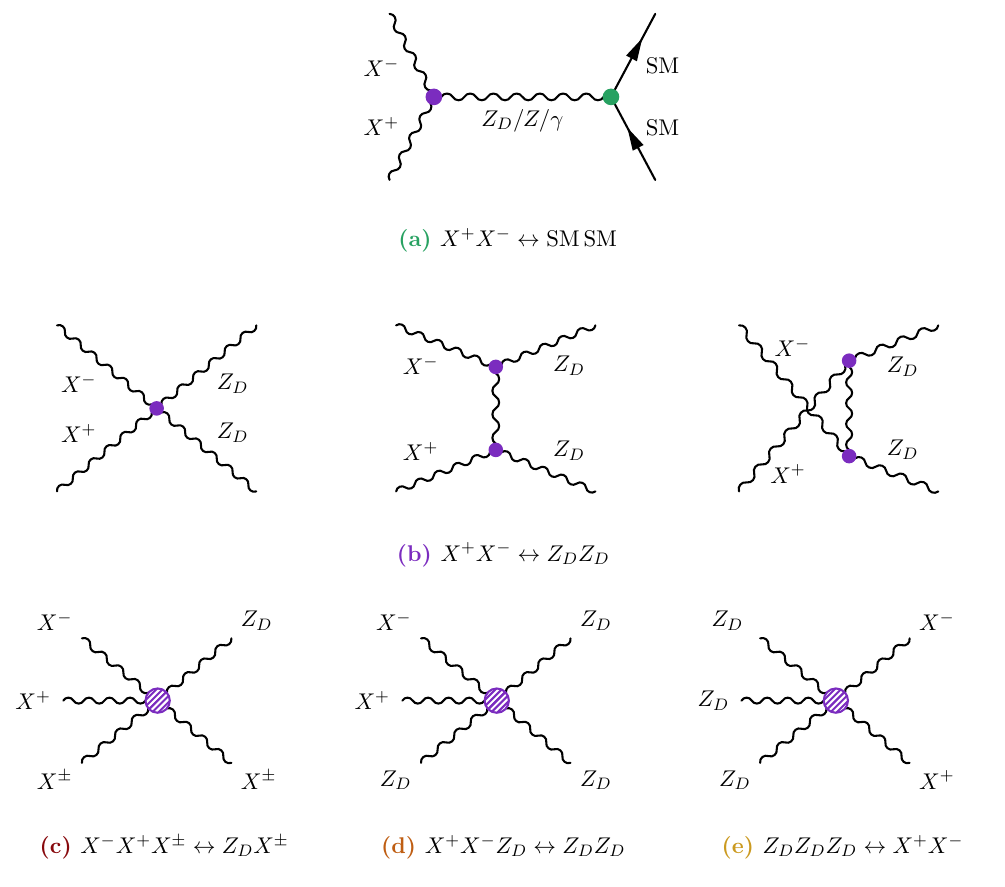}
\end{center}
\vglue -0.8 cm
\caption{\label{fig:BEQdiagrams} Feynman diagrams contributing to the number-changing reactions entering the DM Boltzmann equation: (a) kinetic mixing term, corresponding to an $s$-channel diagram mediated by the dark boson $Z_D$, the $Z$ boson, or the photon $\gamma$ ; (b) dark-sector annihilation via contact, $t$-channel, and $u$-channel diagrams; (c)--(e) three distinct SIMP processes: each bubble stands for several possible tree-level topologies, corresponding to different internal propagators. Vertex colours indicate the relevant couplings: green denotes the kinetic mixing parameter $\epsilon$, and purple the dark gauge coupling $g_X$. }
\label{fig:sidebyside_diagrams}
\end{figure}

The complete Boltzmann equation, including all these processes, is given by 
\be
\begin{aligned}\label{eq:fullbeq}
\dot n_X + 3H n_X = & \, 2 \, \left( 
\gamma_{\scaleto{XX}{4pt}}^{\rm kinetic} 
+ \gamma_{\scaleto{XX}{4pt}}^{\rm dark} 
+ 2   \, \gamma_{\scaleto{XXX}{4pt}} \,  \frac{n_X}{n_X^{\rm eq}} +  \gamma_{\scaleto{XXZ_D}{5pt}} +  \gamma_{\scaleto{Z_D Z_D Z_D}{5pt}} \right) 
\left[1- \frac{n_X^2}{(n_X^{\rm eq})^2}   \right]\,,
\end{aligned}
\ee
where $H$ stands for the Hubble expansion rate, and the general definition for a reaction density $\gamma$, as well as definitions of equilibrium yields and other useful variables are given in appendix~\ref{app:BE}. For readers accustomed to expressing the Boltzmann equation in terms of thermally averaged cross sections, eq.~\eqref{eq:fullbeq} can be equivalently written as
\be
\begin{aligned}\label{eq:fullbeqS}
\dot n_X + 3H n_X  
= & \, 2 \left[1-\frac{n_X^2}{(n_X^{\rm eq})^2 } \right]
\Bigg[ \frac14  \, \ang{\sigma v}_{\scaleto{XX}{4pt}}^{\rm kinetic}  \,  (n_X^{\rm eq})^2 
+ \frac14  \ang{\sigma v}_{\scaleto{XX}{4pt}}^{\rm dark}  \,   (n_X^{\rm eq})^2  \\
 + & \, \frac14  \, \ang{\sigma v^2}_{\scaleto{XXX}{4pt}}  \, n_X \, (n_X^{\rm eq})^2 
 +  \frac14  \, \ang{\sigma v^2}_{\scaleto{XXZ_D}{5pt}} \, n_{Z_D}^{\rm eq} (n_X^{\rm eq})^2 
 + \ang{\sigma v^2}_{\scaleto{Z_D Z_D Z_D}{5pt}} \, (n_{Z_D}^{\rm eq} )^3 \Bigg] \,.
\end{aligned}
\ee
The subscripts in the reaction densities and the cross sections label the following processes\footnote{ We checked that the contributions of the additional channels $X^+ X^- \to Z_D Z$ and $X^+ X^- \to Z Z$ are negligible for the Boltzmann evolution, since they are both suppressed by extra powers of $\sin\alpha$.}:
\begin{equation}\label{eq:processes}
\begin{array}{lcl}
XX \,(\rm{ kinetic}) & : & X^+X^- \leftrightarrow {\rm SM} \, {\rm SM} \\
XX \, (\rm{ dark})     & : & X^+X^- \leftrightarrow Z_D Z_D
\end{array}
\;\;\Bigg|\;\;
\begin{array}{lcl}
XXX        & : & X^-X^+X^\pm \leftrightarrow Z_D X^\pm \\
XXZ_D      & : & X^+X^-Z_D \leftrightarrow Z_D Z_D \\
Z_DZ_DZ_D  & : & Z_D Z_D Z_D \leftrightarrow X^+X^- 
\end{array}
\end{equation}
Here ${\rm SM}$ denotes Standard Model final states (fermions, bosons, and, when relevant, hadrons), while the second column represents the three possible SIMP topologies. The factor of $2$ in the $\gamma_{XXX}$ term of eq.~\eqref{eq:fullbeq} accounts for both channels $X^- X^+ X^{\pm}\leftrightarrow Z_D X^{\pm}$, whereas the other factor arises from the relation $n_{X^+}=n_{X^-}=n_X/2$. All these processes are illustrated by the Feynman diagrams in figure~\ref{fig:BEQdiagrams}. In particular, note that for the process (a), unlike the standard kinetic mixing scenario, there is also a photon contribution, induced by the photon-DM current in eq.~\eqref{eq:Aportal}. Note also that the bubble interactions associated with the SIMP processes actually correspond to $16$, $16$, and $12$ distinct diagrams for the topologies (c), (d), and (e), respectively. To the best of our knowledge, only topology (c) has been considered previously in the literature~\cite{Choi:2019zeb}. In this work, we computed all contributions for the three topologies (including interference effects within each class of diagrams), accounting as well for an arbitrary mass splitting between $X^\pm$ and $Z_D$, using the \texttt{FeynArts}~\cite{Hahn:2000kx} and \texttt{FeynCalc}~\cite{Shtabovenko:2023idz} packages. The final expressions for the cross sections and amplitudes are provided in appendix~\ref{app:BE}.

For the calculation of the reaction densities in eq.~\eqref{eq:fullbeq}, or equivalently, the thermally averaged cross sections in eq.~\eqref{eq:fullbeqS}, we employ the standard Maxwell–Boltzmann approximation. In the case of $2\to2$ processes, this simplifies the phase-space integrals to a one-dimensional integral over the centre-of-mass energy $s$. For $3\to2$ processes, further simplification can be achieved by adopting a non-relativistic approximation, since for temperatures below the DM mass scale the phase space of the initial particles is dominated by small momenta. The details of these approximations, together with the complete expressions for the corresponding cross sections, are collected in appendix~\ref{app:BE}.

Let us now provide some useful details regarding the formulation and solution of the Boltzmann equation for the DM abundance presented above. In general, one must track the abundances of all species that participate in the relevant number-changing processes. For the SM particles, we assume that they remain thermalised throughout the evolution, \emph{i.e.}, they follow equilibrium distributions. In contrast, the dark mediator $Z_D$ thermalises with the plasma only for kinetic mixing values $\epsilon \gtrsim 2 \times 10^{-8}$. For values of $\epsilon$ below this threshold, the evolution of the $Z_D$ abundance, $n_{Z_D}$, must also be tracked through an additional Boltzmann equation. This situation arises, for example, in certain regions of the parameter space relevant for freeze-in production~(see section~\ref{subsec:FI}). For the numerical solution of the Boltzmann equations, we developed our own code based on the publicly available \textsc{ReD-DeLiVeR} package~\cite{Foguel:2024lca,ReD-DeLiVeR}, a Python-based tool for computing relic densities in models with vector mediators.

In what follows, we present and discuss the results of the DM abundance computation for two production mechanisms: thermal freeze-out and non-thermal freeze-in. The first two subsections address the minimal model with $R = 1$ (\textit{i.e.}, $m_{X^\pm} \simeq m_{Z_D}$) for each of these two mechanisms, while the last subsection considers $R > 1$ (as motivated in section~\ref{subsec:split}) in the FO scenario, to illustrate how the dominant process that determines the DM relic abundance can vary with $R$. In particular, a FO regime dominated by 3-to-2 annihilations becomes possible.

\subsection{Freeze-out} \label{subsec:FO}

In the FO mechanism, DM particles are initially in thermal contact with the SM bath, sharing the same temperature $T$. At early times, DM interactions are in equilibrium, but, as the Universe cools below the DM mass, SM-to-DM annihilations become Boltzmann suppressed while DM-to-SM annihilations remain efficient, thus reducing the DM abundance. Eventually, the reaction rate drops below the Hubble expansion rate $H$, marking the moment of chemical decoupling or, equivalently, the freeze-out of the DM species. Afterwards, the DM comoving number density remains essentially constant, setting the relic abundance observed today. Throughout this process, kinetic equilibrium is maintained, meaning that the rate of energy-exchange interactions, typically due to scattering, remains larger than $H$.

\begin{figure}[h!]
\begin{center}
\includegraphics[width=0.9\textwidth]{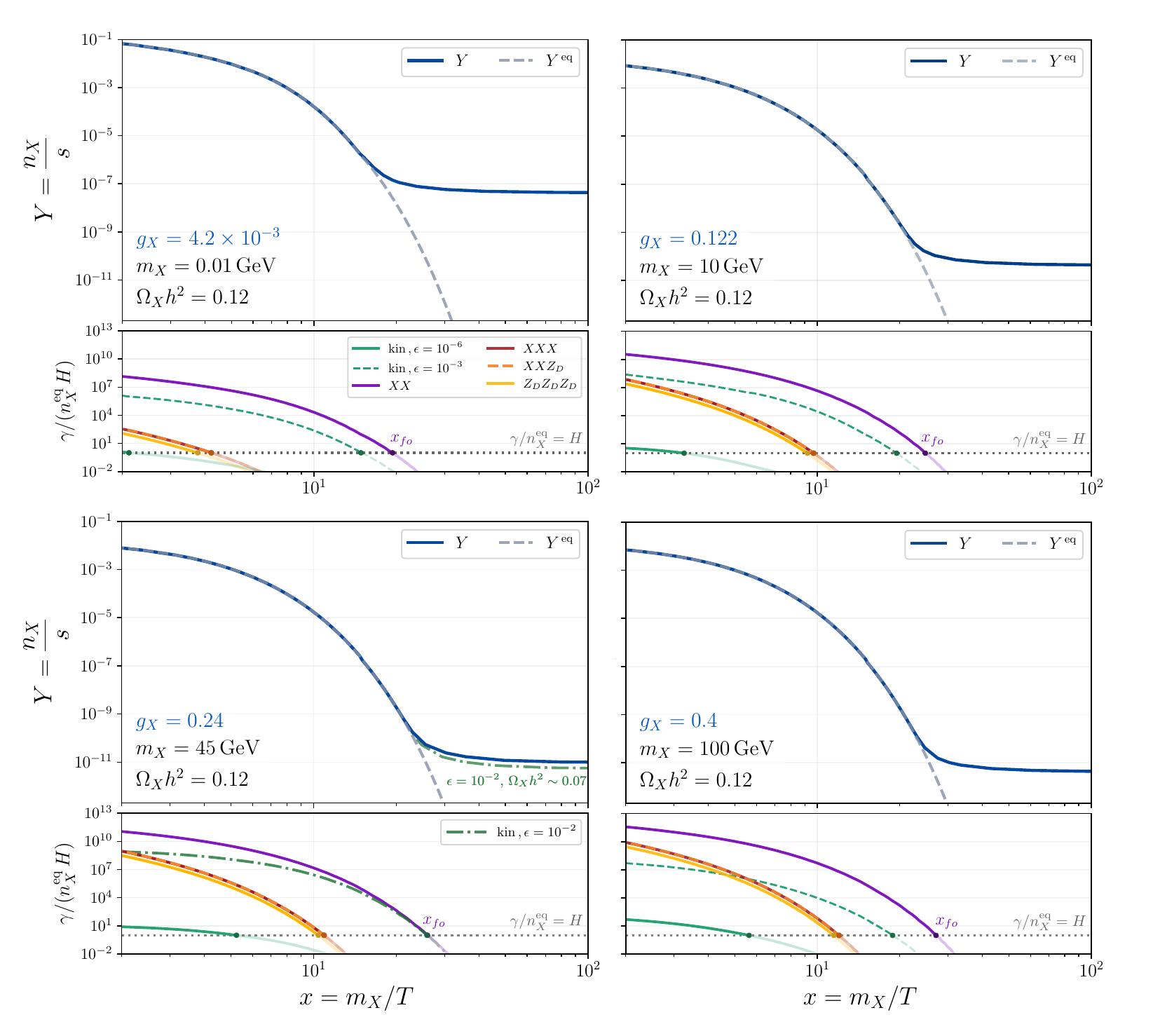}
\end{center}
\vglue -0.3 cm
\caption{\label{fig:FOcases} Evolution of the freeze-out DM yield (upper subpanels) and corresponding rates normalized to the Hubble expansion rate (lower subpanels) for four different DM masses: $m_X/{\rm GeV} =$ 0.01 (top left), 10 (top right), 45
 (bottom left) and $100$ (bottom right). In each plot, the value of $g_X$ is fixed to reproduce the observed DM relic density, $\Omega_X h^2 = 0.12$~\cite{Planck:2018vyg}, for kinetic mixing $\epsilon \lesssim 10^{-3}$, consistent with experimental limits. The solid blue (dashed gray) lines correspond to the comoving (equilibrium) yield $Y$ ($Y_{\rm eq}$) as a function of $x = m_X/T$. The rates are shown for the five relevant channels, as indicated in the legend.
}
\end{figure}

The upper panels of figure~\ref{fig:FOcases} show the DM comoving yield $Y$, defined as $Y\equiv n_X/s$ (where $s$ is the entropy) as a function of the time evolution parameter $x\equiv m_X/T$, obtained by solving eq.~\eqref{eq:fullbeq} for four different DM masses: $m_X/{\rm GeV} =$ 0.01 (top left), 10 (top right), 45 (bottom left) and $100$ (bottom right).\footnote{ Here the mediator mass is essentially the same, $m_{Z_D} \simeq m_X$, since R=1 and corrections are proportional to $\epsilon$.} The gauge coupling $g_X$ value for each panel is chosen in order to reproduce the observed DM relic abundance, $\Omega_X h^2 \simeq 0.12$~\cite{Planck:2018vyg}. The solid (dashed) curve represents the evolution of the DM (equilibrium) yield. The lower panels show the rates, normalized as $\gamma / (n_X^{\rm eq}H)$, for the five processes contributing to eq.~\eqref{eq:fullbeq} and defined by eq.~\eqref{eq:processes}. The horizontal dotted line indicates when the rate equals the Hubble expansion rate, below which the corresponding channel is no longer efficient. One observes that the process that decouples last, and therefore sets the FO temperature, 
$x_{\rm fo}\equiv m_X/T_{\rm fo}$, is the \textbf{dark annihilation}.

For the kinetic-mixing channel, we show the rate for two benchmark values of the mixing parameter: $\epsilon = 10^{-6}$ (solid green) and $10^{-3}$ (dashed green). Note that even for $\epsilon=10^{-3}$ (already almost entirely excluded for most DM masses, as shown in section~\ref{sec:pheno}), the kinetic mixing rate still decouples earlier. To further investigate the impact of this channel for FO, we use the case where $m_X=45$ GeV \textit{i.e.}, near the $Z$-mediated resonance\footnote{Among the three bosons mediating the kinetic-mixing channel, this is the only  possible resonance effect, since in this scenario $m_{Z_D} \simeq m_X$.},  which maximally enhances the channel. We find that only for $\epsilon \gtrsim 10^{-2}$ the kinetic-mixing and dark-annihilation rates become comparable, as shown by the dot-dashed green curve  in the lower left plot of figure~\ref{fig:FOcases}. The upper panel shows the corresponding modification of the relic abundance for $\epsilon=10^{-2}$.

\begin{figure}[tb!]
\begin{center}
\includegraphics[width=0.7\textwidth]{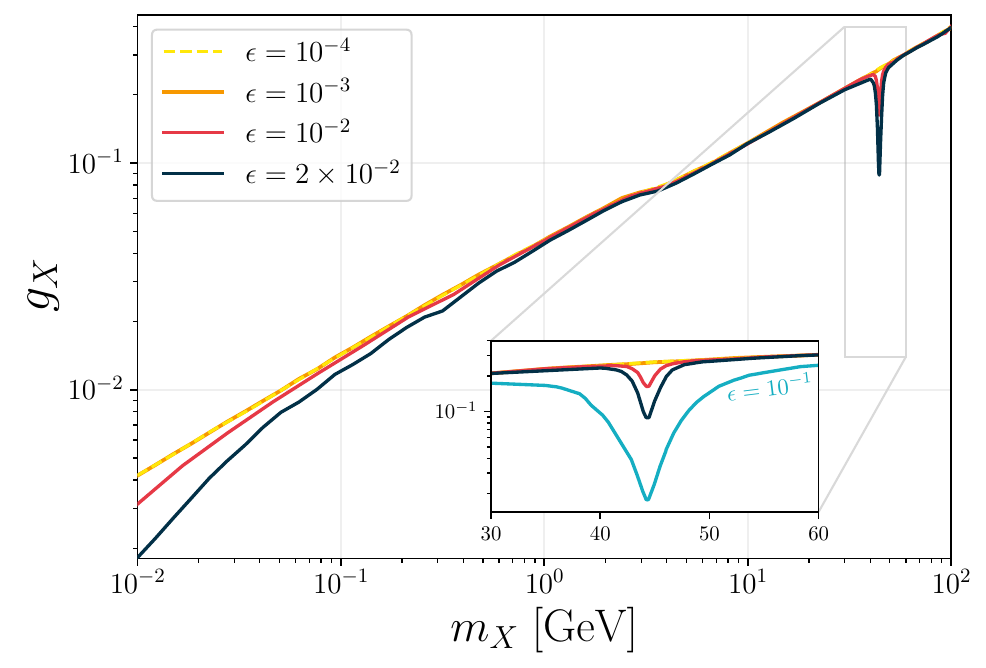}
\end{center}
\vglue -0.5 cm
\caption{\label{fig:ttarget}  Values of the dark gauge coupling $g_X$, as a function of the DM mass $m_X$, which reproduce the observed DM relic abundance, $\Omega_X h^2 = 0.12$~\cite{Planck:2018vyg}. The coloured lines correspond to different values of the kinetic mixing $\epsilon$, as indicated in the legend. The zoomed region near $m_X \simeq 45~\mathrm{GeV}$, corresponding to the $Z$-boson resonance in the $s$-channel, illustrates the enhancement of the kinetic mixing effect. The additional light blue curve in this region represents the case $\epsilon = 10^{-1}$ and is shown only for illustration. }
\end{figure}

Figure~\ref{fig:ttarget} shows the thermal target, i.e.~the collection of values of the parameters $(g_X,\, m_X)$ corresponding to the observed DM density, for different values of the kinetic mixing: $\epsilon=10^{-4}$ (dashed yellow), $10^{-3}$ (solid orange), $10^{-2}$ (solid red), and $2 \times 10^{-2}$ (solid dark blue). Larger values of $\epsilon$ generally reduce the required $g_X$, since the kinetic mixing channel contributes to FO alongside dark annihilation. The zoomed region highlights the enhancement of this effect near $m_X \simeq 45~\mathrm{GeV}$, due to the $Z$-boson $s$-channel resonance of figure~\ref{fig:BEQdiagrams} (a). The light blue curve illustrates the case $\epsilon = 10^{-1}$, which is already experimentally excluded  but is shown here for illustration. Away from the resonance, the kinetic mixing is most relevant for small DM masses, while for $\epsilon \lesssim 10^{-3}$ its impact is negligible across the full DM mass range.
Note that although the precise value of $\epsilon$ is basically irrelevant to set the relic abundance in the FO scenario, the kinetic mixing channel is still crucial to establish thermal contact with the SM sector. As a result, the FO scenario imposes no significant constraint on $\epsilon$, and can therefore be realized for essentially any value of the kinetic-mixing parameter, being limited only by the phenomenological bounds discussed in section~\ref{sec:pheno}.

\subsection{Freeze-in}\label{subsec:FI}

In contrast to the FO mechanism, in the FI scenario, the dark sector particles are never in thermal contact with the SM bath due to the extremely small couplings connecting the two sectors. The initial abundance of the dark species is assumed to vanish, and DM particles are produced via the annihilation of SM states, \textit{i.e.}, through the diagrams of figure~\ref{fig:BEQdiagrams}~(a) considered in the right-to-left direction. A crucial difference is that, while the FO mechanism is insensitive to the initial conditions, in the FI scenario, the onset of DM production depends on the reheating temperature, $T_{\rm rh}$, which is, generically, a free parameter. This temperature is established after the end of inflation, corresponding, for instance, to the decay of the inflaton field into SM particles, and it can vary over several orders of magnitude, from the potentially very large scale of inflation down to the scale of  Big Bang Nucleosynthesis (BBN).\footnote{In fact, to ensure neutrino thermalisation before decoupling one needs $T_{\rm rh} \gtrsim 5$ MeV~\cite{Hannestad:2004px}. The upper limit on $T_{\rm rh}$ depends on the inflationary model, however,  CMB observations bound the tensor-to-scalar perturbation ratio and imposes $ T_{\rm rh }\lesssim 10^{15}$ GeV~\cite{Planck:2018jri}.}

Although reheating marks the onset of freeze-in, the final frozen-in abundance may or may not depend on $T_{\rm rh}$. This is because the FI mechanism can be either \emph{infrared (IR)} dominated or \emph{ultraviolet (UV)} dominated. In the IR regime, most of the DM production occurs at temperatures around the DM mass, so that $T_{\rm rh}$ does not significantly affect the relic density. In the UV regime, instead, the production rate increases with temperature, making the final abundance directly dependent on $T_{\rm rh}$. We will see below that our setup corresponds to the UV-dominated case, with the final DM yield scaling approximately as $T_{\rm rh}^3$. For simplicity, we assume $T_{\rm rh} > v$, thereby avoiding complications associated with electro-weak phase transition (EWPT) (see, however, Refs.~\cite{Baker:2017zwx,Bian:2018mkl,Benso:2025vgm} for studies of this and other possible dark-sector phase transitions). Therefore, we consider the SM degrees of freedom to be those of the electroweak-unbroken phase.

Furthermore, we also fix an upper bound on the reheating temperature by requiring $T_{\rm rh} < v_X$, so that FI production occurs entirely in the $SU(2)_X$ broken phase. This condition automatically guarantees the validity of the effective field theory description. Indeed, from eq.~\eqref{KM} one has schematically $\Lambda \sim v_X/\sqrt{\epsilon}$ after symmetry breaking. Working in the regime $\epsilon \ll 1$, it follows that $\Lambda > v_X$, and therefore the hierarchy $\Lambda > v_X > T_{\rm rh}$ is satisfied.

Besides the DM species, the dark mediator is also produced via the kinetic mixing portal. At first sight, one might expect that the dominant contribution comes from inverse decays, \textit{i.e.}, the coalescence of two fermions into a $Z_D$. However, even though we always work in the regime before the EWPT, fermions are not massless due to temperature effects: their effective thermal masses scale as the temperature times the gauge coupling divided by a loop factor. As a result, at high temperatures, $T_{\rm rh} \gg m_X \sim m_{Z_D}$, inverse decays are kinematically forbidden. Instead, the dark mediator is populated through $t$-channel processes such as $f \bar f \to Z_D \Phi$, $f \bar f \to Z_D V$ with $V = W^i, B$, and $q \bar q \to Z_D \, G$ with $G$ the gluon.

\begin{figure}[t!]
\begin{center}
\includegraphics[width=0.6\textwidth]{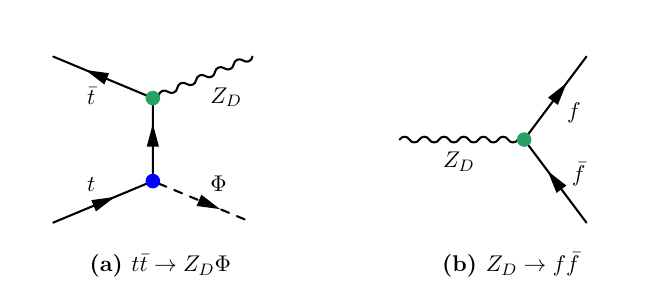}
\end{center}
\vglue -0.8 cm
\caption{\label{fig:dZDbeq} Diagrams relevant for the $Z_D$ Boltzmann equation: (a) top anti-top annihilation into a dark mediator $Z_D$ and a Higgs doublet $\Phi$ ; (b) $Z_D$ decay into two fermions. Green vertices represent kinetic mixing insertions, while blue stands for the top Yukawa couplings.}
\end{figure}

It is important to track the dark mediator abundance because, depending on its evolution, the number density $n_{Z_D}$ may eventually become large enough to produce DM, through the processes in the second and third columns of figure~\ref{fig:BEQdiagrams}. Therefore, to correctly compute the DM abundance in the FI scenario, we solve the system of two coupled Boltzmann equations for the number densities of the DM particles, $n_X$, and of the dark mediator, $n_{Z_D}$, given by
\be
\begin{aligned}\label{eq:FIbeqDM}
\dot n_X + 3H n_X = \, & 2 \, 
\gamma_{\scaleto{XX}{4pt}}^{\rm kinetic} \, \left[1- \frac{n_X^2}{(n_X^{\rm eq})^2}   \right] + 2 \mathcal{T}_{\scaleto{XX}{4pt}}^{\, \rm dark} + 2 \, \mathcal{T}_{\scaleto{XXX}{4pt}}  + 2 \, \mathcal{T}_{\scaleto{XXZ_D}{5pt}} +2 \,  \mathcal{T}_{\scaleto{Z_DZ_DZ_D}{5pt}} \, ,
\end{aligned}
\ee
\vspace{10pt}
\be
\begin{aligned}\label{eq:FIbeqZD}
\dot n_{Z_D} + 3H n_{Z_D}  =  \, - \left(2 \mathcal{T}_{\scaleto{XX}{4pt}}^{\, \rm dark} +  \, \mathcal{T}_{\scaleto{XXX}{4pt}}  +  \, \mathcal{T}_{\scaleto{XXZ_D}{5pt}} + 3 \,  \mathcal{T}_{\scaleto{Z_DZ_DZ_D}{5pt}} \right) + \left( \gamma_{Z_D}  + \gamma_{t} \right) \, \left[1- \frac{n_{Z_D}}{n_{Z_D}^{\rm eq}}   \right] \,,
\end{aligned}
\ee
where we defined 
\be
\begin{aligned}\label{eq:FIbeqT}
& \mathcal{T}_{\scaleto{XX}{4pt}}^{\, \rm dark} \equiv    \, \gamma_{\scaleto{XX}{4pt}}^{\rm dark} \, \left[ \frac{n_{Z_D}^2}{(n_{Z_D}^{\rm eq})^2} - \frac{n_X^2}{(n_X^{\rm eq})^2}   \right] \, , \quad \quad 
\mathcal{T}_{\scaleto{XXX}{4pt}}   \equiv   \, 2 \, \gamma_{\scaleto{XXX}{4pt}}  \,  \frac{n_X}{n_X^{\rm eq}} \, \left[ 
\frac{n_{Z_D}}{n_{Z_D}^{\rm eq}}
- \frac{n_X^2}{(n_X^{\rm eq})^2}   \right] \,,\\
& \mathcal{T}_{\scaleto{XXZ_D}{5pt}}
\equiv  \, \gamma_{\scaleto{XXZ_D}{5pt}} \,  \frac{n_{Z_D}}{n_{Z_D}^{\rm eq}} \, \left[ \frac{n_{Z_D}}{n_{Z_D}^{\rm eq}} - \frac{n_X^2}{(n_X^{\rm eq})^2}   \right] \, , \quad \quad  \mathcal{T}_{\scaleto{Z_DZ_DZ_D}{5pt}} \equiv  \,  \gamma_{\scaleto{Z_D Z_D Z_D}{5pt}}  \, \left[ \frac{n_{Z_D}^3}{(n_{Z_D}^{\rm eq})^3} - \frac{n_X^2}{(n_X^{\rm eq})^2}   \right] \, , 
\end{aligned}
\ee
which replace the corresponding terms of the previous DM Boltzmann eq.~\eqref{eq:fullbeq}, since now $n_{Z_D} \neq n_{Z_D}^{\rm eq}$. 
In the dark-mediator Boltzmann eq.~\eqref{eq:FIbeqZD}
we added two $X$-independent source terms: mediator decays, with 
$\gamma_{Z_D} \equiv \langle \Gamma_{Z_D \to f \bar f} \rangle \, n_{Z_D}^{\rm eq}$, and mediator production via $t$-channel top-quark annihilation\footnote{For the final computation we only included the top channel, as it scales with the large top Yukawa coupling (see, for instance, Ref.~\cite{Caputo:2018zky} for similar considerations in a different context). We checked that channels involving gauge bosons provide similar contributions but have no impact on the final DM relic density result.}, with
$\gamma_{t} \equiv \langle \sigma v \rangle_{t t \to Z_D \Phi} \, (n_{t}^{\rm eq})^2$, both illustrated in figure~\ref{fig:dZDbeq}.

Before presenting the results, let us stress that, in our vector-DM scenario with kinetic mixing induced by a dimension-6 operator, the FI production is UV dominated. The reason is that the amplitude of the dominant channel, $f \bar f \to X^+ X^-$, grows with energy since the longitudinal polarisations of the external vector-DM states scale as $\epsilon^\mu_L \sim p^\mu/m_X$. An analogous behaviour occurs in the SM for fermion annihilation into longitudinal $W$ bosons, where the growth with energy is cancelled by the Higgs contribution. In our case the growth with $s$ continues up to the cutoff of the effective theory (where it will be eventually cured by a given UV completion). As a consequence, the production rate is dominated by the highest available temperature, \textit{i.e.},~the DM abundance is saturated near the reheating temperature $T_{\rm rh}$, and it does not change significantly during the subsequent evolution. Note that the main consequence of the UV sensitivity is the introduction of this additional parameter, the reheating temperature $T_{\rm rh}$, which also controls the DM abundance. While this enlarges the region of parameter space compatible with the observed relic density, it also implies that the abundance depends on the post-inflationary cosmological history rather than solely on particle-physics parameters.

\begin{figure}[ht]
\begin{center}
\vspace{3pt}
\includegraphics[width=0.8\textwidth]{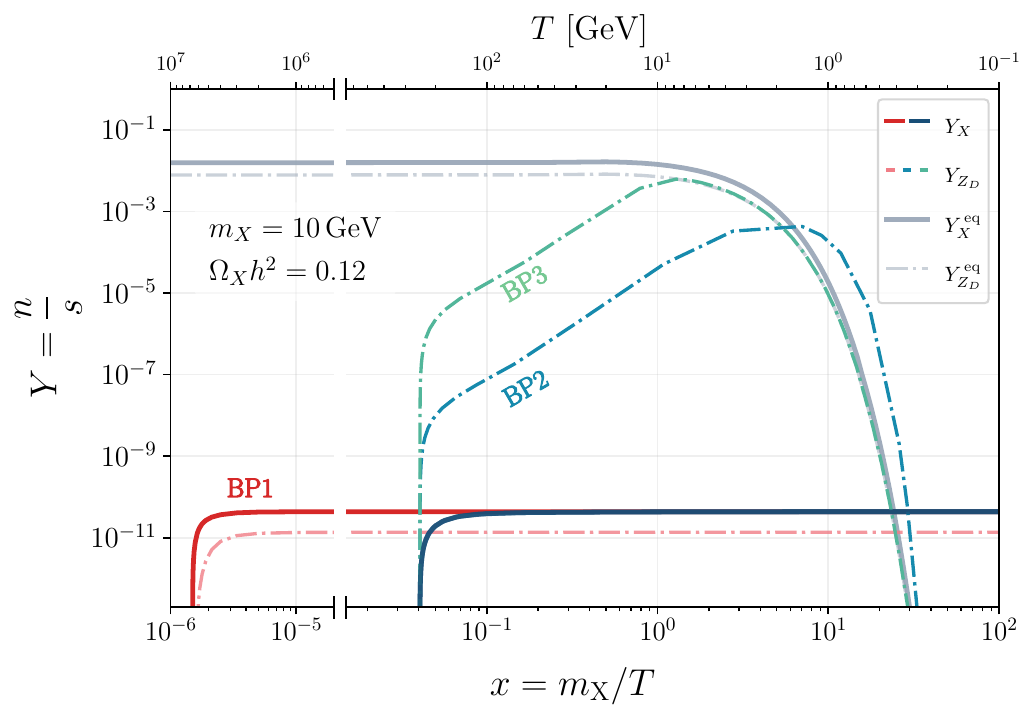}
\end{center}
\vglue -0.6 cm
\caption{\label{fig:FIcurves} Evolution of the freeze-in DM and $Z_{D}$ yields as a function of $x$ for $m_{Z_D} \simeq m_X = 10~\mathrm{GeV}$. Solid lines represent the DM yield, dot-dashed lines the dark-mediator yield, and gray curves the corresponding equilibrium yields. The colors distinguish between the three different benchmark scenarios defined by fixed $(T_{\rm rh}, g_X, \epsilon)$ values: \textbf{BP1} = $(v_X,\,3 \times 10^{-6},\,2.5 \times 10^{-15})$ in red, \textbf{BP2} = $(v,\,3 \times 10^{-6},\,5.4 \times 10^{-9})$ in blue, and \textbf{BP3} = $(v,\,2 \times 10^{-7},\,8.15 \times 10^{-8})$ in green, all of which reproduce the observed DM relic abundance, $\Omega_X h^2 =0.12$. }
\end{figure}

To illustrate the behaviour of the FI solutions we have selected three Benchmark Points (BP). In figure~\ref{fig:FIcurves} we show the evolution of the species yield $Y$ as a function of $x$ for these points and $m_X= 10$ GeV, for both the DM candidate $X$ (solid lines) and for the dark mediator $Z_D$ (dot-dashed lines). Gray curves represent the corresponding equilibrium yields, and the upper x-axis indicates the corresponding temperature. Since the relevant temperature range spans several orders of magnitude, the x-axes have a cut. The initial abundance of dark species at $T_{\rm rh}$ is always set to zero.\footnote{In principle, an initial population could be generated via gravitational particle production. However, for the reheating temperatures relevant to our setup such contributions are negligible~\cite{Kolb:2023ydq}.}

The first point, \textbf{BP1} is fixed at $T_{\rm rh} = v_X$ and $(g_X,\epsilon) = (3 \times 10^{-6},\,2.5 \times 10^{-15})$. The DM yield $Y_X$ is shown in solid red, while the dark mediator yield $Y_{Z_D}$ is shown in dot-dashed light red. Due to the UV-dominated nature of FI, the DM abundance saturates at early times. The dark mediator exhibits the same early saturation, since at such high temperatures the dark-annihilation diagrams in figure~\ref{fig:BEQdiagrams} (b) (which also grow with energy) have a relatively large rate (though still below the Hubble one), dominating over other $Z_D$ production channels. Consequently, once the DM abundance saturates, it efficiently drives the production of $Z_D$. This behaviour changes at lower $T_{\rm rh}$, where the dark-annihilation contribution becomes subdominant with respect to top-quark annihilation, which then controls $Z_D$ production.

To show that dynamics, we selected a lower temperature for the points \textbf{BP2} (blue) and \textbf{BP3} (green), both corresponding to $T_{\rm rh} = v$, with $(g_X,\epsilon) = (3 \times 10^{-6},\,5.4 \times 10^{-9})$ and $(g_X,\epsilon) = (2 \times 10^{-7},\,8.15 \times 10^{-8})$, respectively. In these cases, the DM abundance again saturates near $T \sim T_{\rm rh}$, while the dark mediator continues to be produced until top-pair annihilation becomes Boltzmann suppressed. Subsequently, $Z_D$ decays into SM states, depleting  $Y_{Z_D}$. Note that this depletion will always occur, although for \textbf{BP1} it is not shown, since the mediator’s longer lifetime makes its decay efficient only at temperatures below the range displayed in figure~\ref{fig:FIcurves}. The crucial difference between \textbf{BP2} and \textbf{BP3} resides in the size of $\epsilon$. For \textbf{BP2}, $\epsilon$ is smaller than the thermalisation threshold, so $Y_{Z_D}$ never follows the equilibrium curve. In contrast, for \textbf{BP3} the larger value of $\epsilon$ allows $Y_{Z_D}$ to reach and follow its equilibrium distribution, as shown by the dot-dashed green curve.

\begin{figure}[b!]
\begin{center}
\includegraphics[width=0.82\textwidth]{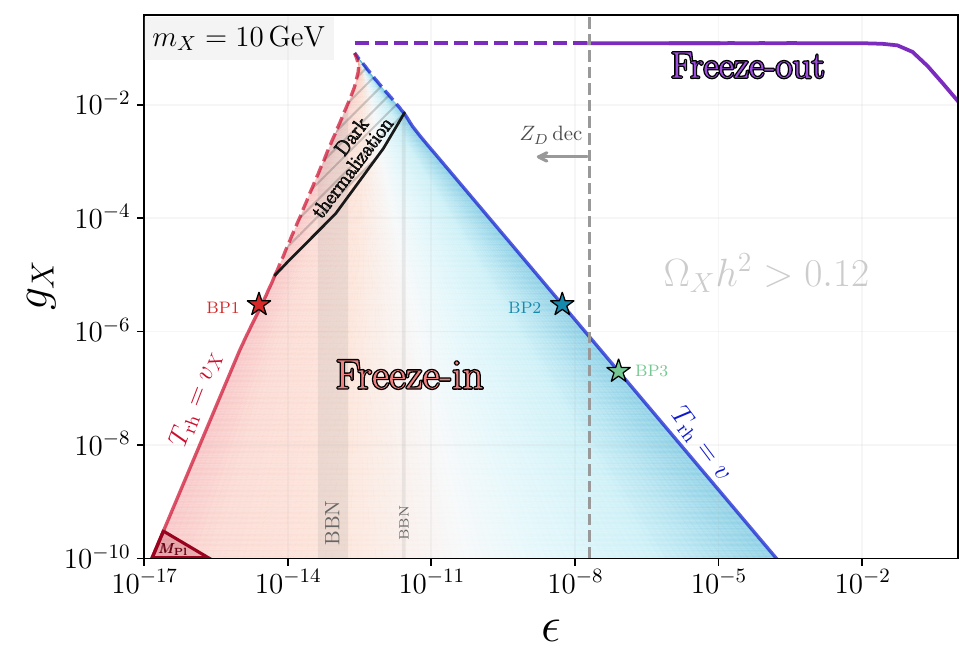}
\end{center}
\vglue -1 cm
\caption{\label{fig:FImesa} {\em Phase diagram} that summarizes the regions where FO and FI  can achieve the observed DM relic abundance in the $(g_X , \epsilon)$ plane for a reference value $m_X = 10$ GeV. The purple line corresponds to freeze-out, while the red and blue lines define the boundaries of the freeze-in region, for $T_{\rm rh} = v_X$ and $T_{\rm rh} = v$, respectively, with intermediate reheating temperatures lying in between. The dashed gray line at $\epsilon = 2 \times 10^{-8}$ marks the thermalisation threshold below which the dark mediator $Z_D$ is decoupled from the SM. The dashed lines/regions in both the FO and FI cases indicate dark sector thermalisation, \textit{i.e.}, points where a dark temperature $T_X$ should be considered. The dark red region in the lower left corner corresponds to $\Lambda > M_{\rm Pl}$. Light-gray vertical bands show the regions excluded by BBN~\cite{Fradette:2014sza}. The three benchmark points, indicated by coloured stars along the FI boundaries, are the ones already discussed and shown in figure~\ref{fig:FIcurves}.}
\end{figure}

We synthesize our findings for FO and FI in the {\em phase diagram} of figure~\ref{fig:FImesa}. There we show the different DM production regimes in the $(g_X ,\epsilon)$ plane for fixed $m_{Z_D} \simeq m_X = 10~\mathrm{GeV}$ (this representation is inspired by Refs.~\cite{Chu:2011be,Hambye:2019dwd}). Let us go through the different regions in the plot. 

Starting from the right-hand side, the solid purple line indicates the values of $g_X$ and $\epsilon$ that yield the correct freeze-out DM relic abundance. This line is almost horizontal, since $g_X$ sets the relic density through dark annihilation, as discussed previously, and only for very large values of $\epsilon$ the kinetic mixing channel can play a compensating role. Moving to the left, one eventually crosses the threshold $\epsilon \sim 2 \times 10^{-8}$, below which the dark sector no longer thermalises with the SM. Beyond this point, the purple line becomes dashed, corresponding to a secluded FO scenario, where DM freezes out in the dark sector at a temperature $T_X$,  different from that of the SM bath.

Moving downwards to smaller values of $g_X$, the triangular region corresponds to a successful freeze-in DM production. Along the red line one reproduces the observed relic density for a reheating temperature $T_{\rm rh} = v_X$, while the blue line corresponds to $T_{\rm rh} = v$. For intermediate values, $v_X < T_{\rm rh} < v$, one can also reproduce the correct abundance for couplings lying between the red and blue curves. In the upper part of the triangle, the dashed region corresponds to dark-sector thermalisation, \textit{i.e.}, the region where scatterings within dark-sector particles are in equilibrium, and where one should, in principle, define and track an independent dark temperature, $T_X$. To determine the threshold (black solid line) marking the onset of the dark-sector thermalisation, we computed the scattering rate for $X^+ X^- \leftrightarrow Z_D Z_D$.\footnote{Note that self-scattering among the $X^{\pm}$ particles could also lead to thermalisation within the dark sector. The rate of this process becomes efficient in a region similar to that for the $X^\pm-Z_D$ scattering.} Then, for each fixed value of $\epsilon$, we scanned the $(g_X, T_{\rm rh})$ parameter space to identify the points where (i) $\Omega_X h^2 = 0.12$ is obtained, and (ii) such scattering rate reaches the Hubble rate at some moment in time. In this way, we ensure that for smaller $g_X$ the system remains in the non-thermalised regime, where our result is accurate. We emphasise that the dashed region should be taken with a grain of salt, as in this work we did not solve the full Boltzmann equations accounting for different temperatures and therefore corrections are expected (see~\cite{Bernal:2020gzm} for an estimate of the correction factor and~\cite{Chu:2011be} for a full treatment).

In the lower left corner of figure \ref{fig:FImesa} there is also a small region corresponding to $\Lambda >M_{\rm Pl}$, where the effective approach breaks down. The vertical light gray bands represent the regions excluded by BBN~\cite{Fradette:2014sza}, \textit{i.e.}, for these $\epsilon$ values the mediator would decay during BBN and spoil its predictions. Finally, the region between the blue and purple curves leads to DM overproduction.

\subsection{Comparing annihilation rates as a function of $R = m_{Z_D}/m_X$}
\label{subsec:FOsplit}

We have seen in subsection~\ref{subsec:FO} that the kinetic-mixing channel is not relevant for setting the FO relic abundance when $R = 1$, corresponding to our minimal model. Variations of this setup with $R > 1$ were discussed in section~\ref{subsec:split}. In these scenarios, it is possible to enhance the contribution of the kinetic-mixing $s$-channel diagram of figure~\ref{fig:BEQdiagrams}(a) by taking larger values of $R$, thereby approaching the $Z_D$ resonance at $R=2$. Alternatively, when $\epsilon$ is relatively small, we will see that the regime $R>1$ can shift the dominant mechanism to SIMP processes.

\begin{figure}[tb!]
\begin{center}
\vspace{10pt}
\includegraphics[width=0.48\textwidth]{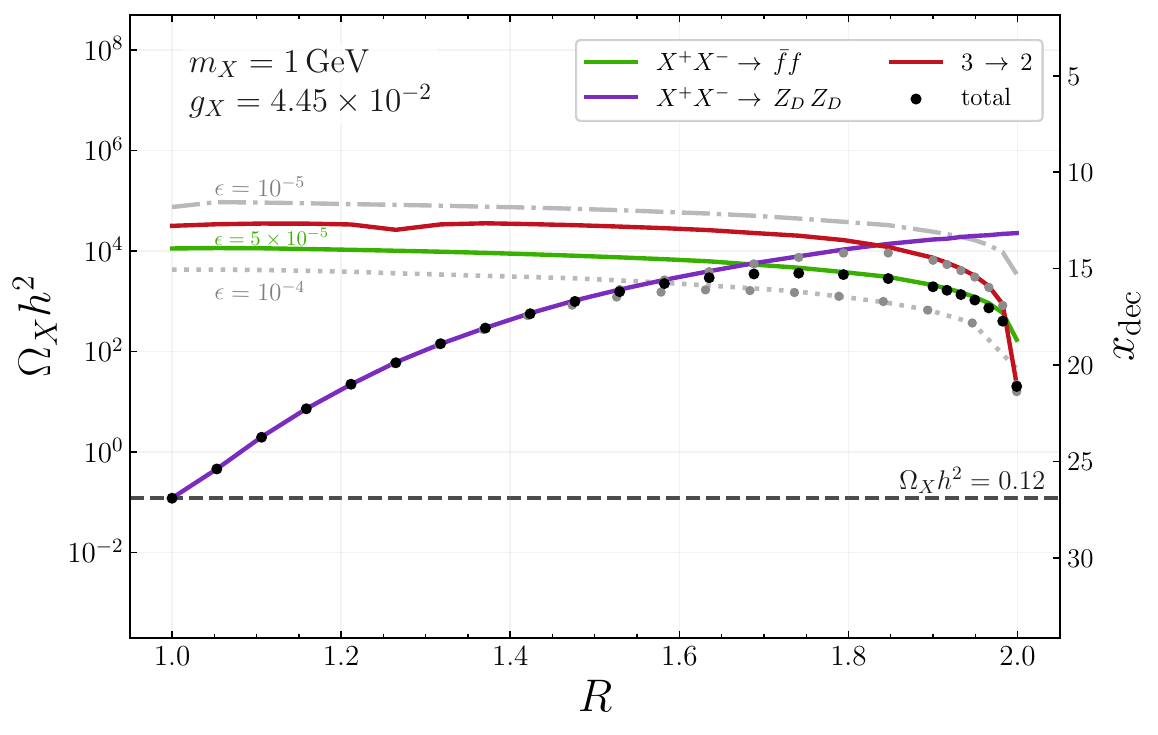}
\includegraphics[width=0.48\textwidth]{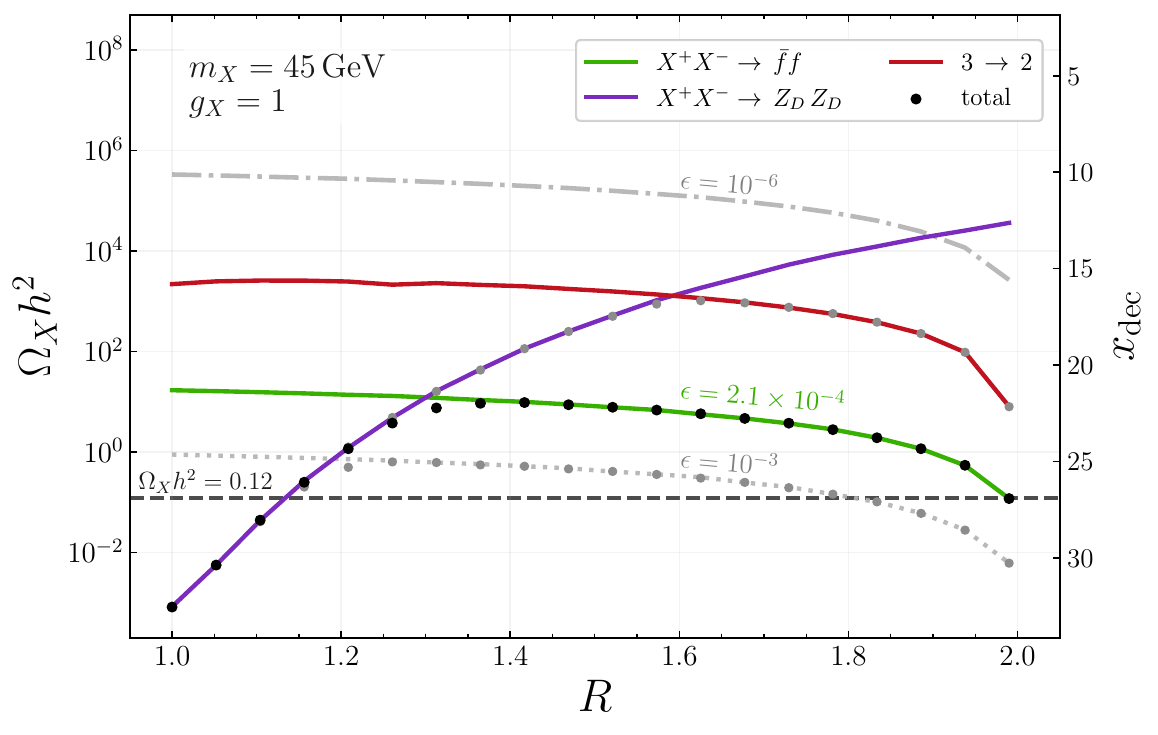}
\includegraphics[width=0.48\textwidth]{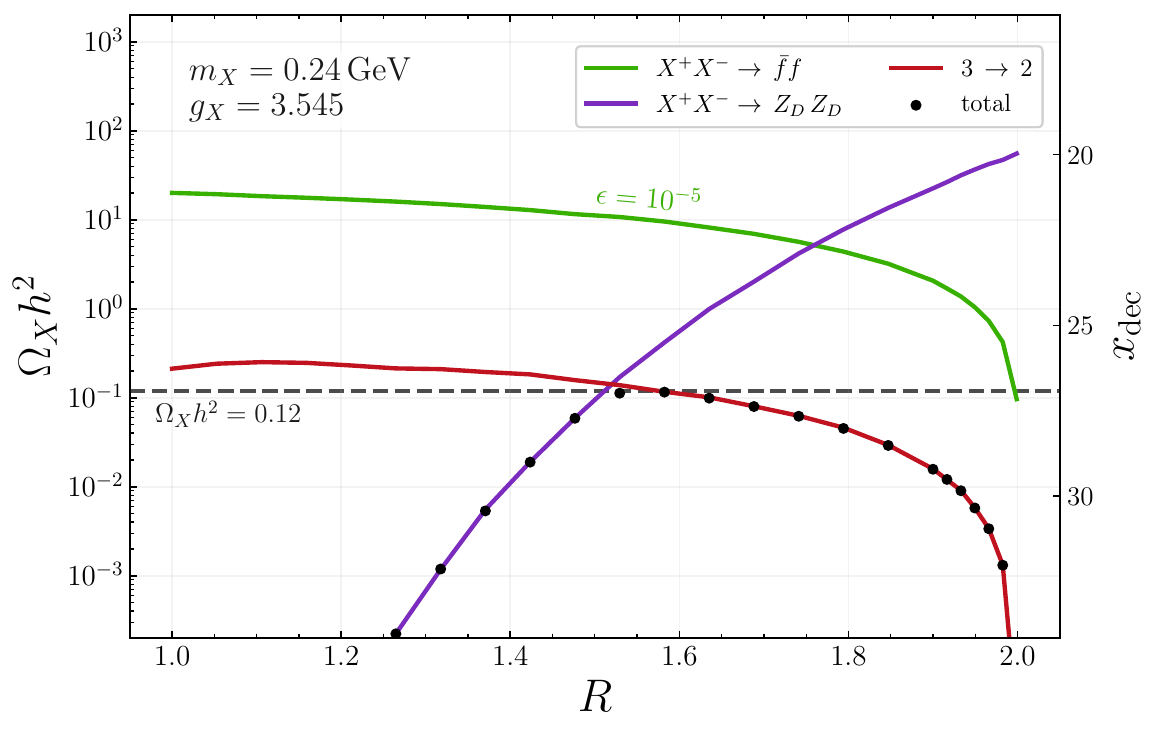}
\end{center}
\vglue -0.5 cm
\caption{\label{fig:omegavsR} Relic density $\Omega_X h^2$ as a function of the mass splitting $R \equiv m_{Z_D}/m_X$, obtained by considering each channel in isolation: kinetic mixing $X^+ X^- \to \bar f f$ (green), dark annihilation $X^+ X^- \to Z_D Z_D$ (purple), and SIMP $3 \to 2$ (red). Black dots indicate the final relic abundance when all channels are included simultaneously, while the horizontal dashed black line corresponds to $\Omega_X h^2 = 0.12$. Each panel corresponds to a fixed choice of $m_X$ and $g_X$, as indicated. Additional gray dotted and dot-dashed curves show different choices of $\epsilon$, to illustrate how the dominant channel (the lowest curve) varies with this parameter.}
\end{figure}

In figure~\ref{fig:omegavsR} we show the predicted DM relic density as a function of the mass-splitting parameter $R$. The three panels correspond to different choices of $(m_X, g_X)$, while the curves indicate the relic abundance obtained by solving the Boltzmann equation with only one channel switched on at a time: kinetic mixing $X^+ X^- \to \bar f f$ (green), dark annihilation $X^+ X^- \to Z_D Z_D$ (purple), and SIMP $3 \to 2$ (red). Channels that alone would predict larger values of $\Omega_X h^2$ are less efficient, since DM would decouple earlier from the SM bath. The channel that actually determines the relic density is the one yielding the smallest $\Omega_X h^2$, corresponding to the latest chemical decoupling. The right $y$-axis of each panel translates the relic abundance into the corresponding decoupling epoch, defined by $x_{\rm dec} = m_X/T_{\rm dec}$.  The black dots indicate the final relic density, obtained when all channels are  simultaneously included, while the horizontal dashed black line stands for the observed relic density. Note that the kinetic mixing only affects the green curve, decreasing (increasing) the relic abundance for larger (smaller) values of $\epsilon$.

In the upper left panel we take $m_X = 1~\mathrm{GeV}$ and $g_X = 4.45 \times 10^{-2}$, which reproduces the observed relic abundance for $R=1$ through dark annihilation. This corresponds to the behaviour already discussed in section~\ref{subsec:FO}. As $R$ increases, dark annihilation becomes less efficient, and the dominant mechanism eventually shifts either to kinetic mixing or to the SIMP processes, depending on the value of the kinetic-mixing parameter $\epsilon$. The green curve corresponds to $\epsilon = 5 \times 10^{-5}$, for which the transition to the kinetic-mixing dominance occurs around $R \simeq 1.7$. For larger $\epsilon$, such as $\epsilon = 10^{-4}$ (gray dotted), the transition takes place at smaller $R$, whereas for smaller values, e.g., $\epsilon = 10^{-5}$ (gray dot-dashed), the dark-annihilation regime switches to the SIMP one around $R \simeq 1.8$.

A similar behaviour is also observed in the other panels of figure~\ref{fig:omegavsR}. In the upper right panel we fix $m_X = 45~\mathrm{GeV}$ and $g_X = 1$, corresponding to the case where kinetic mixing is enhanced by the $Z$-boson resonance. For $\epsilon = 2.1 \times 10^{-4}$, the observed relic density is reproduced through kinetic mixing at $R = 2$. For the same mass and coupling, the correct relic density can also be obtained via dark annihilation around $R \simeq 1.15$. Once again, for larger kinetic mixing ($\epsilon = 10^{-3}$, dot-dashed) the regime transition occurs earlier, whereas for very small values ($\epsilon = 10^{-6}$, dotted) the SIMP channel comes to dominate for $R \gtrsim 1.6$. In the lower panel we consider a scenario with a larger gauge coupling, which further enhances the SIMP channels, by fixing $m_X = 0.24~\mathrm{GeV}$ and $g_X = 3.545$. In this case, the $3 \to 2$ regime already dominates for $R \gtrsim 1.5$, yielding the observed relic abundance around $R\simeq 1.55$.

\begin{figure}[t!]
\begin{center}
\includegraphics[width=0.49\textwidth]{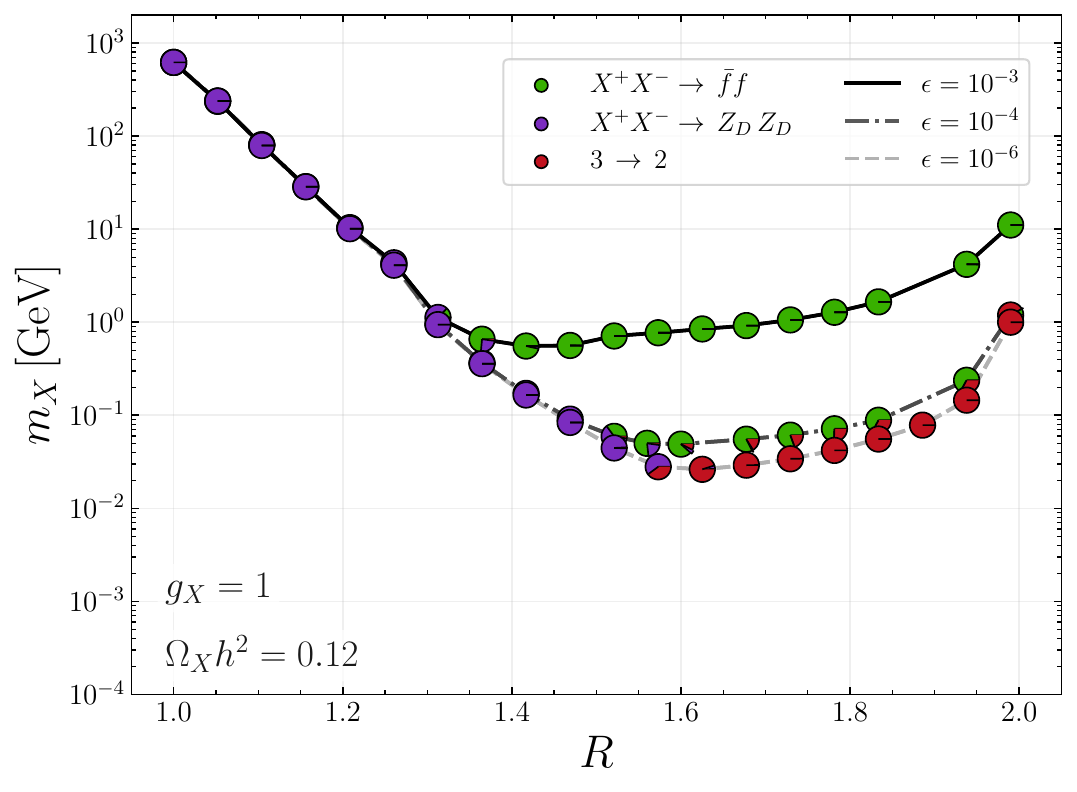}
\includegraphics[width=0.49\textwidth]{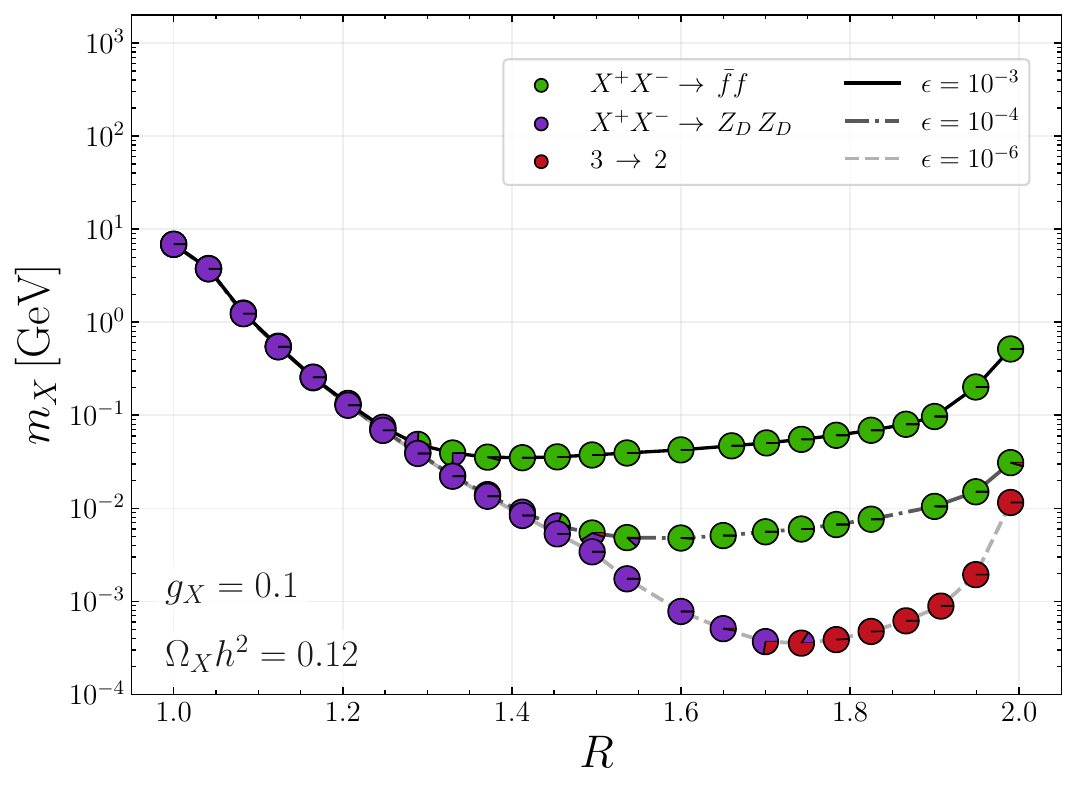}
\end{center}
\vglue -0.5 cm
\caption{\label{fig:mXvsR} Curves of $m_X$ as a function of $R$, which reproduce the observed relic abundance $\Omega_X h^2 = 0.12$, for $g_X = 1$ (left panel) and 0.1 (right panel), and for $\epsilon = 10^{-3}$ (solid), $10^{-4}$ (dot-dashed), \rm{\, and \,} $10^{-6}$ (dashed). Selected points along each curve are illustrated with pie charts, which show the relative contribution of the three regimes to the relic density: kinetic mixing (green), dark annihilation (purple), and SIMP (red).}
\end{figure}

Figure~\ref{fig:mXvsR} provides a complementary perspective on the regime transitions as the mass splitting increases by showing the curves in the $(m_X,\,R)$ plane that reproduce the observed DM relic value $\Omega_X h^2 = 0.12$. Selected points along the curves are illustrated with pie charts, which display the relative contribution of each channel to the relic abundance, using the same colour code as in figure~\ref{fig:omegavsR}. The two panels correspond to fixed gauge couplings, $g_X = 1$ (left) and 0.1 (right), while the three lines correspond to different kinetic mixing values: $\epsilon = 10^{-3}$ (solid), $10^{-4}$ (dot-dashed), and $10^{-6}$ (dashed). For small $R$, dark annihilation dominates, as shown by the fully purple pie charts, and the value of $\epsilon$ has no impact. As the mass splitting increases, the dominant channel flips either to the kinetic mixing regime (for larger $\epsilon$) or to the SIMP regime (for smaller $\epsilon$). In the latter case, further decreasing $\epsilon$ has little effect, since the dynamics are already fully dominated by the dark gauge-coupling  diagrams. In the transition region, the pie charts display mixed colors, reflecting simultaneous contributions from two or even three channels. Finally, the comparison between the two panels shows that for smaller $g_X$ the curves shift downward toward lighter DM masses, in order to reproduce the observed relic abundance.

%

\section{Phenomenology of the vector dark matter model}
\label{sec:pheno}

The non-abelian kinetic mixing,  induced by the dimension-6 operator in eq.~\eqref{KM}, opens new avenues to probe the $SU(2)_D$ dark sector through the kinetic mixing portal. The dark mediator can be produced in various experimental settings, including colliders and beam-dump experiments, as well as in astrophysical environments. In addition, cosmological bounds, such as those from the Cosmic Microwave Background radiation (CMB) and the BBN, play an important role in constraining the parameter space at very small values of $\epsilon$.

In this section we first describe in detail the different experimental searches and production mechanisms that can constrain the kinetic mixing portal. We then compile these constraints in the model parameter space, and compare them with the target curves and regions which reproduce the observed DM relic density. Finally, we comment on additional search strategies, including DM direct and indirect detection. Let us note that recently a study investigating the phenomenological consequences of a closely related framework with vector dark matter and magnetic-dipole interactions has appeared~\cite{Banerjee:2025ejz}. We refer the reader to Ref.~\cite{Banerjee:2025ejz} for additional constraints from direct detection and beam-dump experiments in this similar setup.

\vspace{-0.3cm}
\subsection{Bounds on the vector mediator $Z_D$}

\begin{figure}[tb!]
\begin{center}
\includegraphics[width=0.9\textwidth]{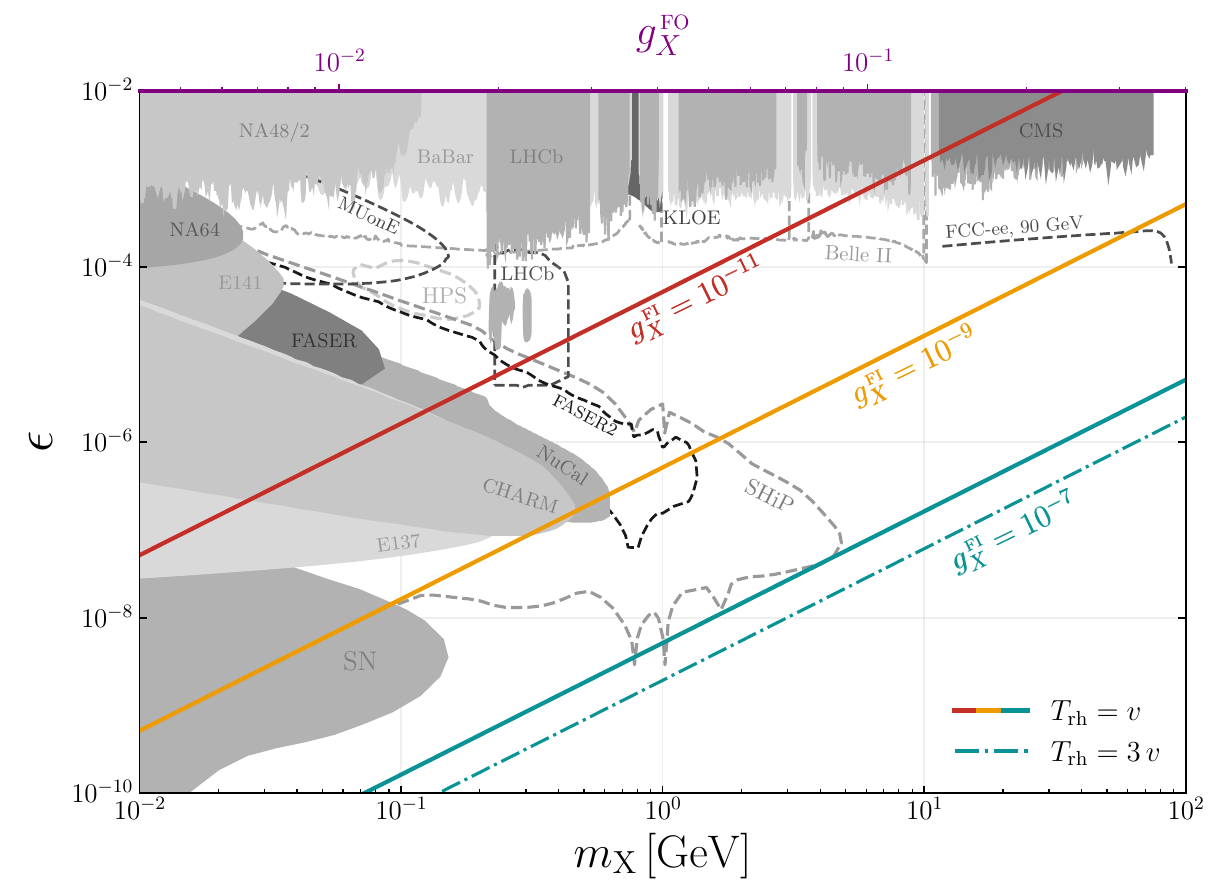}
\end{center}
\vglue -0.3 cm
\caption{\label{fig:bounds} Constraints on the kinetic mixing $\epsilon$ as a function of the DM mass $m_X$, with the dark photon mass $m_{Z_D} \simeq m_X$. Gray shaded regions indicate bounds from current experimental searches and astrophysical limits, while dashed lines show future projections. See text for technical details and references regarding each bound. Coloured solid lines correspond to the parameter values yielding the observed DM abundance in the freeze-in regime for different gauge couplings (as indicated by labels), assuming a fixed reheating temperature $T_{\rm rh} = v$. The blue dash-dotted line illustrates the case $T_{\rm rh} = 3 \, v$. The upper x-axis shows the gauge coupling $g_X^{\rm FO}$ reproducing the observed DM relic density in the freeze-out regime for the corresponding $m_X$, independently from the value of $\epsilon$ (except for $\epsilon \gg 10^{-3}$, which, however, is already strongly constrained by experiments). }  
\end{figure}

The most relevant current and projected searches for a massive dark photon $Z_D$ can be grouped into distinct classes, including both experimental searches and astrophysical/cosmological constraints, as outlined below. The corresponding bounds are summarized in figure~\ref{fig:bounds}, which compiles all relevant limits in the $(\epsilon, m_X)$ plane. We focus on our minimal model, where $m_{Z_D} \simeq m_X$. For extensive reviews and compilations of vector mediator/dark photon bounds, see Refs.~\cite{Fabbrichesi:2020wbt,Batell:2022dpx,Antel:2023hkf}.

\begin{enumerate}[label=\roman*.]
    \item \textbf{hadron colliders:} benefiting from high centre-of-mass energies and luminosities, these facilities can probe the largest mediator masses. The most constraining bounds currently come from the LHC experiments. The \textbf{LHCb} detector is sensitive to dark mediator production through several channels, including meson decays, proton-proton collisions ($p\, p \to p\, p\, Z_D$), and Drell–Yan annihilation ($q \bar q \to Z_D$) at higher energies. The mediator subsequently decays into $e^+e^-$ or $\mu^+\mu^-$ pairs, with searches classified as prompt or displaced depending on the decay length. In addition to current bounds~\cite{LHCb:2019vmc}, LHCb projections extend down to $\epsilon \sim 10^{-5}$ for mediator masses $0.2 \lesssim m_X/{\rm GeV} \lesssim 0.4$~\cite{PhysRevLett.116.251803}. The \textbf{CMS} experiment provides the leading bounds at higher masses, $m_X \gtrsim 20~{\rm GeV}$, where production is dominated by Drell–Yan and detection proceeds via decays into leptonic final states~\cite{CMS:2019kiy}. 

    Beyond the main LHC detectors, forward experiments along the beam collision axis place stringent limits on long-lived mediators. The \textbf{FASER} experiment, located about $480$~m downstream from the ATLAS interaction point, began data-taking in 2022 and currently sets the strongest bounds in the range $20 \lesssim m_X/{\rm MeV} \lesssim 100$ for $10^{-5} \lesssim \epsilon \lesssim 10^{-4}$~\cite{FASER:2023tle}. Its proposed upgrade, \textbf{FASER2}, is expected to substantially extend the sensitivity region during the HL-LHC era~\cite{Kling:2021fwx,Anchordoqui:2021ghd}.

    \item \textbf{$e^+e^-$ colliders:} with lower backgrounds compared to hadron colliders, these experiments typically provide cleaner signatures. The dominant production mechanism is $e^+e^- \to \gamma \, Z_D$, followed by $Z_D \to \ell^+ \ell^-$. The strongest current bounds from such dilepton resonance searches come from the \textbf{BaBar} experiment at the PEP-II asymmetric-energy $e^+e^-$ $B$-factory at SLAC~\cite{BaBar:2001yhh}, which collected data between 1999 and 2008 at a center-of-mass energy $\sqrt{s} \simeq 10$~GeV. Looking ahead, the \textbf{Belle II} detector~\cite{Belle-II:2018jsg}, located at SuperKEKB in Japan, is expected to provide the most stringent limits in the mass range $0.4 \lesssim m_X/{\rm GeV} \lesssim 10$, probing the upper part of the $(\epsilon,\, m_X)$ plane.

     Besides these two, the \textbf{KLOE} experiment at the DA$\Phi$NE $e^+e^-$ collider in Frascati has also performed dedicated searches for dark photons in the sub-GeV mass range. Operating at $\sqrt{s} \simeq 1.02$~GeV, corresponding to the $\phi$-meson resonance, KLOE is sensitive to mediator production through $\phi \to \eta Z_D$ and radiative processes $e^+e^- \to \gamma Z_D$. The resulting limits are obtained from a combination of $Z_D \to \mu^+\mu^-$ and $Z_D \to \pi^+\pi^-$ decay channels, yielding stringent bounds in the range $0.1 \lesssim m_X/{\rm GeV} \lesssim 1$, with sensitivity down to $\epsilon \sim 10^{-3}$–$10^{-4}$~\cite{Anastasi:2018azp}.

    In the longer term, the proposed \textbf{FCC-ee} (Future Circular Collider – electron–positron) at CERN represents the next generation of $e^+e^-$ colliders. Preliminary projections indicate sensitivity down to $\epsilon \sim 10^{-4}$ for mediator masses in the range $10$–$100$~GeV, assuming operation at $\sqrt{s} = 90$~GeV~\cite{Karliner:2015tga}.

    \item \textbf{$\ell$ beam-dump:} in this type of search, a high-intensity lepton beam is dumped on a fixed target, where the vector mediator can be produced through processes such as lepton Bremsstrahlung ($\ell N \to \ell N Z_D$, with $N$ the target nucleus) and subsequently decay into visible SM particles downstream. The most relevant bounds in this category come from the E137, E141, and NA64 experiments. Both \textbf{E137}~\cite{Bjorken:1988as,Andreas:2012mt} and \textbf{E141}~\cite{Riordan:1987aw}, operated at SLAC in the 1980s with a 20~GeV electron beam, provided constraints that remain world-leading in their respective mass ranges. The \textbf{NA64} experiment, located at the CERN Super Proton Synchrotron (SPS), is an active $e^-$ beam dump operating with a $100$–$150$~GeV electron beam since 2016. The bounds derived in~\cite{NA64:2019auh} probe the region $10 \lesssim m_X/{\rm MeV} \lesssim 25$ for kinetic mixing values in the range $\epsilon \sim 10^{-3}$–$10^{-4}$.

    Besides, the on-going \textbf{HPS} (Heavy Photon Search) experiment has already published results from prompt searches~\cite{HPS:2018xkw}. HPS employs a high-intensity electron beam on a thin tungsten target to probe both prompt and displaced $Z_D \to e^+e^-$ decays in the $20$–$500$~MeV mass range. Its projected displaced-vertex sensitivity will explore the currently unconstrained region around $\epsilon \sim 10^{-4}$~\cite{Baltzell:2016eee,Baltzell:2022rpd}. The upcoming \textbf{MUonE} experiment~\cite{MUonE:2019qlm} will also probe this parameter space, being sensitive to mediator masses up to $\sim 100$~MeV and mixings in the range $10^{-3} \lesssim \epsilon \lesssim 10^{-4}$ by exploiting a $160$~GeV muon beam from the CERN M2 line~\cite{Rocha:2025rqc}.

    \item \textbf{$p$ beam-dump:} the main difference here is the use of a high-intensity proton beam. When dumped on a fixed target, it produces a large number of secondary mesons ($M=\pi^0, \eta, K, D, \ldots$), which can decay into a photon and the dark mediator ($M \to \gamma Z_D$). As in hadron colliders, additional production channels include proton Bremsstrahlung ($p N \to p N Z_D$, with $N$ the target nucleus) and Drell–Yan processes at higher energies. After being produced, the mediator propagates and decays downstream into visible SM particles inside the detector. Relevant bounds have been obtained by experiments such as the \textbf{CHARM} experiment~\cite{Bergsma:1985qz}, which operated at the CERN SPS in the 1980s with a 400~GeV proton beam on a copper target; the \textbf{NuCal} experiment~\cite{Blumlein:1991xh}, which ran at Protvino in the late 1980s with a 70~GeV proton beam on an iron target; and the \textbf{NA48/2} experiment~\cite{Batley:2015lha}, which mainly collected data at the CERN SPS in the 2000s with a 400~GeV proton beam on a beryllium target, for which the relevant bound comes from a prompt dark-photon search.

    In the future, the approved \textbf{SHiP} (Search for Hidden Particles) experiment~\cite{SHiP:2025ows} at CERN will employ the SPS 400~GeV proton beam in a setup explicitly optimized for long-lived particle searches, in contrast to the previously mentioned beam-dump experiments, which were originally designed as general-purpose neutrino detectors. As a result, SHiP will significantly extend the sensitivity to the kinetic mixing portal.

    \item \textbf{astrophysics:} the most relevant bound in the parameter space considered here comes from observations of Supernova (SN) 1987A~\cite{Chang:2016ntp}. The idea is that, if vector mediators were produced during the SN explosion, they would contribute, in addition to neutrinos, to the energy-loss budget. The standard constraint requires that the instantaneous luminosity released into new particles does not exceed the Raffelt criterion, $L_{\, \rm new} \lesssim 3 \times 10^{52}\,{\rm erg/s}$~\cite{Raffelt:1996wa}.

    \item \textbf{cosmology:} limits typically arise from CMB and BBN considerations~\cite{Fradette:2014sza}, but these apply only for very small kinetic mixing values, $\epsilon \lesssim 10^{-10}$, and are therefore not shown in figure~\ref{fig:bounds}.

\end{enumerate}

Figure~\ref{fig:bounds} compares all the bounds discussed above with the requirement to reproduce the observed DM abundance.
This allows to confront the theoretically motivated regions of parameters with the experimental sensitivities. On the one hand, the diagonal coloured lines correspond to the FI production regime, showing the values of $\epsilon$ and $m_X$ that yield the correct relic density, for gauge coupling $g_X^{\rm FI} = 10^{-11}$ (red), $10^{-9}$ (yellow), and $10^{-7}$ (blue), assuming a reheating temperature $T_{\rm rh} = v$. Points below the lines underproduce DM, while points above overproduce it. Increasing the reheating temperature $T_{\rm rh}$ shifts the corresponding freeze-in target curve downward, towards smaller $\epsilon$. As an illustration, the case $g_X^{\rm FI} = 10^{-7}$ and $T_{\rm rh} = 3\,v$ is shown with a blue dash-dotted line. On the other hand, in the FO regime, for each DM mass $m_X$ there exists a gauge coupling $g_X$ that reproduces the DM abundance, essentially independently from $\epsilon$ (a small dependence occurs only for $\epsilon \gg 10^{-3}$, see figure~\ref{fig:ttarget}). The upper x-axis of the plot indicates such value of $g_X^{\rm FO}$, corresponding to the $m_X$-dependence shown by the orange line in figure~\ref{fig:ttarget}.

Let us also note that figure~\ref{fig:bounds} is plotted as a function of the DM mass $m_X$, and corresponds to our minimal case $R = 1$. For the extended scenarios with $1 < R < 2$, the phenomenological bounds (gray regions) keep the same shape but simply shift along the $m_X$ axis, according to $m_X = m_{Z_D}/R$. For $R> 2$, the situation is different since the mediator can decay invisibly into DM states, thus modifying the constraints. On the other hand, the DM abundance curves in the figure would be significantly modified when $R>1$.  

\subsection{Bounds on the vector dark matter $X^\pm$}

Having described the bounds that directly constrain the production of the kinetic mixing mediator $Z_D$, we now turn to probes of the DM candidates $X^\pm$. 
In addition to the usual portals through the $Z_D$ and the $Z$ boson (enhanced when $m_{Z_D} \gtrsim 10 \,\text{GeV}$), the same dimension-6 operator that generates the kinetic mixing also induces an effective photon portal (see section \ref{sec:mixings}), given by
\begin{equation}
\mathcal{L}_{XA} \;=\; i \, g_X \, \epsilon \, A_\mu \,
\partial_\nu \!\left(X^{-\nu} X^{+\mu} - X^{+\nu} X^{-\mu}\right) ~,
\end{equation}
which constitutes a specific feature of the present scenario. Before turning to more generic bounds on the DM, let us first discuss those associated with this photon portal.

As already discussed in section \ref{sec:mixings}, this interaction does not give our DM candidate a millicharge, since it does not have the Lorentz structure of a minimal gauge coupling and therefore it does not generate a long-range Coulomb potential. Indeed, to obtain the potential for e.g.~DM--fermion scattering mediated by a photon, one can compute the scattering amplitude, take the non-relativistic (NR) limit, and Fourier-transform it to position space. Here the amplitude carries one explicit factor of DM momentum from the derivative, in addition to the momentum factors from the longitudinal polarisations. Consequently, in the non-relativistic limit the amplitude vanishes and no long-range potential arises.\footnote{ For comparison, in the case of the SM $W^\pm$ bosons, gauge invariance ensures the presence of additional interaction terms (such as $A_\nu W^+_\mu \partial^\nu W^{-\mu}$), where the derivative is contracted with the off-shell photon rather than with a massive vector. This structure enables the recovery of the long-range Coulomb interaction, as expected for electrically charged particles.} Therefore, the usual limits on millicharged particles (see, e.g., Ref.~\cite{Cirelli:2024ssz} for a review) do not apply in our case.\footnote{ For related works involving millicharged sub-GeV dark matter with massive dark photons, see e.g. Ref.~\cite{Feng:2023ubl,Cheung:2007ut,Feldman:2007wj}} 

Let us briefly discuss why usual bounds on millicharged DM are evaded. They are typically divided into three categories: cosmology, astrophysics, and direct detection. 
\textbf{Cosmology:} CMB constraints arise because truly millicharged DM remains tightly coupled to the baryon-photon plasma at recombination, altering the acoustic peaks~\cite{Dolgov:2013una}. In our case, however, the scattering amplitudes vanish in the NR limit, as explained above, so our DM candidates are not tightly coupled to the plasma, evading these bounds. 
\textbf{Astrophysics:} Galactic dynamics constraints rely on the existence of long-range interactions between DM and the interstellar medium or magnetic fields, which could lead to phenomena such as spin-down of the galactic disk or the expulsion of millicharged particles from the halo~\cite{Stebbins:2019xjr,Chuzhoy:2008zy}. Since our interaction is short-ranged, such effects are absent. 
\textbf{Direct detection:} In terrestrial experiments, such as XENON~\cite{Essig:2012yx,Essig:2017kqs} and CDMS-II~\cite{CDMS-II:2009ktb}, the usual millicharge bounds arise from standard fermionic millicharged currents, which generate Coulomb interactions. In our case, the effective operator instead induces momentum- (or velocity-) suppressed couplings, dramatically reducing the sensitivity of existing searches. In addition, our vector DM carries spin one, corresponding to a different operator basis relevant for scattering on SM targets~\cite{Catena:2019hzw}.

A comprehensive assessment of DM direct detection constraints is left for future work, as it requires a dedicated analysis including both the $Z_D$-mediated and the effective photon (dipole-like) interactions, and a careful evaluation of their respective contributions to DM--nucleon and DM--electron scattering. Related studies of multipole DM-photon interactions can be found in Refs.~\cite{Hisano:2020cmy,Chu:2023xxx}, which set interesting bounds on such scenarios. Nevertheless, such constraints are not expected to significantly affect the scenario presented in figure~\ref{fig:bounds}, since in our case the DM scatterings with SM targets are always proportional to the product $g_X \cdot \epsilon$. In the freeze-in regime, this product is extremely small, and for increasing reheating temperature $T_{\rm rh}$ it can be further suppressed, keeping direct detection signals well below current and projected sensitivities. In the freeze-out scenario, as the correct DM abundance does not depend on $\epsilon$, it always can be taken sufficiently small to evade the bounds. 

To conclude this section, let us briefly comment on \textbf{indirect detection} and \textbf{self-interactions}. For the former, the thermally-averaged DM annihilation cross section into SM states, diagram in figure~\ref{fig:BEQdiagrams} (a), is $p$-wave suppressed due to the derivative nature of the vector coupling, \textit{i.e.}~it scales with the DM velocity, which is extremely small today. As a result, indirect detection bounds can be evaded. Moreover, in scenarios where the relic density is set by FI production, such limits are negligible due to the tiny couplings involved. 

Coming to self-interactions, since our DM candidates are spin-one bosons from a non-abelian symmetry, they feature $2 \to 2$ self-scattering processes such as $X^{+} X^{-} \to X^{+} X^{-}$ and $X^{\pm} X^{\pm} \to X^{\pm} X^{\pm}$. The corresponding self-interaction cross section is constrained by astrophysical observations, including the Bullet Cluster~\cite{Markevitch:2003at,Clowe:2003tk,Randall:2008ppe,Robertson:2016xjh} and the shapes or ellipticities of halos~\cite{Peter:2012jh,Kahlhoefer:2013dca}, to satisfy $\sigma_{\rm self}/m_X \lesssim 1 \, {\rm cm^2/g}$. We have estimated the relevant cross sections from these processes, which scale as $\sigma_{\rm self} \sim g_{X}^{4}$, and verified that saturating the observational limit would require relatively large couplings, $g_X \gtrsim 1$. 
Therefore, the parameter space considered in this work is consistent with self-interaction bounds.

\section{Conclusions} \label{sec:conclusion}

In recent years, the exploration of GeV and sub-GeV dark sectors has gained increasing importance, driven by advances in precision data and the particle physics community efforts at the so-called intensity frontier.
In the related quest for DM, without observational clues on its nature, one must rely on some guiding principles to build motivated models. Among these, symmetry considerations and the pursuit of predictivity have historically played a central role. 

In this work we focused on the vector portal, but departing from the canonical approach where the SM is extended by an additional $U(1)_X$ gauge symmetry. Instead, we considered a dark sector with an $SU(2)_X$ gauge symmetry, and with the kinetic mixing arising via a dimension-six operator. This setup leads to a distinctive pattern: a custodial $SU(2)_D$ symmetry enforces mass degeneracy among the three gauge bosons, and its unbroken subgroup $U(1)_D$ ensures the stability of the vectors $X^\pm$ carrying a dark charge, which thus serve as DM candidates. At the same time, the neutral dark vector $Z_D$ acts as a mediator through the kinetic-mixing portal. In addition, the same higher-dimensional operator induces an effective DM-photon dipole interaction. The construction is remarkably minimal, being fully specified by just three free parameters: the common dark-vector mass $m_X$, the dark gauge coupling $g_X$, and the kinetic mixing coefficient $\epsilon$.

Given these ingredients, we demonstrated that the observed DM relic abundance can be achieved either by freeze-out or freeze-in over a wide range of parameters. We highlighted how different DM number-changing processes may dominate in different regimes, ranging from kinetic-mixing annihilations to dark-sector annihilations, either 2-to-2 or 3-to-2 (also known as SIMP).
 
In the FO case, although the kinetic mixing is useful to maintain thermal equilibrium between the dark sector and the SM bath, the final relic density is fully determined by a dark-annihilation process, $X^+ X^- \to Z_D Z_D$, and thus depends solely on $g_X$ and $m_X$. 
Conversely, in the FI case, production is dominated by the kinetic mixing, through SM annihilations into DM states. Due to the effective nature of the dimension-six operator, the corresponding amplitude grows with energy. Therefore, the FI mechanism is UV-dominated, and the relic abundance is extremely sensitive to the reheating temperature $T_{\rm rh}$. Depending on the value of $g_X$, the production of dark mediators can become efficient enough for dark annihilations to also affect the DM abundance. For this reason, we solved the two coupled Boltzmann equations governing the yields of both $X^\pm$ and $Z_D$. Figure~\ref{fig:FImesa} serves as the main summary of the production mechanisms, highlighting the regions, in the plane of couplings, where the observed DM abundance can be successfully reproduced, $\Omega_X h^2 = 0.12$. 

Variations of our minimal construction allow for 
a non-negligible mass splitting between 
$m_{Z_D}$ and $m_{X}$, which we parametrize by their ratio $R$.
We find that, as $R$ increases,  the dominant annihilation channel in the FO scenario can shift from 2-to-2 dark annihilations to 
SIMP processes (when $\epsilon$ is sufficiently small) or to kinetic-mixing processes 
(for larger $\epsilon$ values). We note in passing that our analysis of
the 3-to-2 processes among vector bosons is the most complete to date.

Finally, we explored the observational constraints on the model parameters. Since the dark vector mediator is a massive dark photon, one can apply essentially all experimental bounds associated with a dark-photon production and subsequent decay into SM states, as well as astrophysical and cosmological limits. These constraints impose restrictions
on the $(\epsilon, m_{X})$ plane, as the equality $m_{Z_D}\simeq m_X$ holds in our minimal model. We contrasted a compilation of such bounds with the regions that reproduce the correct DM abundance, as illustrated in figure~\ref{fig:bounds}. DM indirect detection and self-interaction constraints can be safely neglected, while DM direct detection bounds would require a dedicated study, taking into account that our spin-one DM candidate scatters on SM targets through a set of specific velocity-suppressed interactions. We leave a quantitative study of the direct detection phenomenology for future work, but we already demonstrated that DM direct detection bounds can be qualitatively evaded in a large, viable region of the model parameters.

In summary, our scenario unifies the vector DM and mediator properties within a compact, predictive model, with a wide testable parameter space, providing a well-motivated target for the future exploration of the dark sector.

\section*{Acknowledgements}

We would like to thank Thomas Hambye, Felix Kahlhoefer and Enrique Fernandez-Martinez for valuable comments and suggestions. MF thanks the IF-USP (Brazil) for the warm hospitality when this work was started. MF has received support from the European Union Horizon 2020 research and innovation program under the Marie Sk\l odowska-Curie grant agreement No 101086085–ASYMMETRY. ALF is supported by Funda\c{c}\~ao de Amparo \`a Pesquisa do Estado de S\~ao Paulo (FAPESP) under the contracts 2022/04263-5, and 2024/06544-7. ALF thanks the Institute of Theoretical Physics IFT-UAM (Madrid) for hospitality while this work was being completed. R.Z.F. is partially supported by FAPESP under contract No. 2019/04837-9, and by Conselho Nacional de Desenvolvimento Científico e Tecnológico (CNPq).

\appendix
\renewcommand{\theequation}{\Alph{section}\arabic{equation}}
\makeatletter
\@addtoreset{equation}{section}
\makeatother

\section{Details on the model parameters and interactions}
\label{details}

Let us study the precise relations between the kinetic mixing $\epsilon$ introduced in eq.~\eqref{KM}, the vector boson mass parameters $m_X^2$ and $\hat m^2_Z$ introduced in eq.~\eqref{Lm}, the $Z-Z_D$ mixing angle $\alpha$ introduced in eq.~\eqref{AZ}, and the physical vector boson masses $m_Z$ and $m_{Z_D}$.

Since $\theta_w$ and $m_Z$ are precisely measured, there are only two free physical parameters, $\epsilon$ and $r\equiv m^2_{Z_D}/m^2_Z$. Experimental bounds require $\epsilon\ll 1$, and our study mostly focuses on the region $r\ll 1$, even though the following equations apply to any value of $r$.
These two free parameters determine the mixing angle $\alpha$, as well as the unmixed mass parameters $m^2_X$ and $\hat m^2_Z$, as follows:
\be 
\sin 2\alpha = t_w \epsilon \frac{(1-\epsilon^2/c_w^2)^{1/2}(1+r)+D}
{(1-\epsilon^2)(1-r)}~,
\qquad 
\cos 2\alpha = \frac{-t_w^2 \epsilon^2 (1+r) + (1-\epsilon^2/c_w^2)^{1/2} D}
{(1-\epsilon^2)(1-r)}
~,
\label{alpha}\ee
\be 
\hat m_Z^2 = \frac{(1-\epsilon^2/c_w^2)(1+r)+(1-\epsilon^2/c_w^2)^{1/2}D}
{2(1-\epsilon^2)}m_Z^2~,
\quad
m_X^2 = \frac{(1-\epsilon^2/c_w^2) (1+r)
-(1-\epsilon^2/c_w^2)^{1/2} D}{2}m_Z^2
~,
\label{mu}\ee
where 
\be
D\equiv \left[(1-\epsilon^2)(1-r)^2-t_w^2\epsilon^2(1+r)^2\right]^{1/2}~,
\ee
Note that $D$ vanishes for $r = r_\pm$, where
\be
r_\pm =\dfrac{(1-\epsilon)^{1/2}\pm t_w \epsilon}{(1-\epsilon)^{1/2}\mp t_w \epsilon} \,. 
\ee
At $r=r_\pm$ the mixing is close to maximal, $\tan 2\alpha = -(1-\epsilon^2/c_{w}^2)^{1/2}/(t_w\epsilon)$, 
and one has $\hat m_Z^2=m_X^2$.
For a fixed $\epsilon$, $D(r)$ is a decreasing function of $r$ from $D(0)=(1-\epsilon^2/c_w^2)^{1/2}$ to $D(r_-)=0$: in this region $m^2_{Z_D} < m_Z^2$. For $r_-<r<r_+$ there is no solution, that is, in this model $Z_D$ and $Z$ cannot be exactly degenerate in mass. Finally,  
$D(r)$ is increasing from $D(r_+)=0$ to $D(r)\simeq r(1-\epsilon^2/c_w^2)^{1/2}$ for $r\gg 1$: in this region $m^2_{Z_D} > m_Z^2$.

Expanding eqs.~\eqref{alpha} and \eqref{mu} in $\epsilon$, one obtains 
\be 
\sin 2\alpha = \frac{2t_w}{1-r} \epsilon +{\cal O}(\epsilon^3) ~,
\qquad 
\cos 2\alpha = 1 -\frac{2t_w^2}{(1-r)^2} \epsilon^2 +{\cal O}(\epsilon^4) ~,
\ee
\be 
\hat m_Z^2 = \left[1-\frac{t_w^2 \epsilon^2}{1-r} +{\cal O}(\epsilon^4)\right]m_Z^2 ~,
\qquad
m_X^2 = \left[1-\frac{\epsilon^2}{c_w^2} + \frac{t_w^2\epsilon^2}{1-r}+{\cal O}(\epsilon^4)\right] m_{Z_D}^2 ~.
\ee
While $\hat m_Z^2$ is proportional to $m_Z^2$, note that $m_X^2$ is proportional to $m_{Z_D}^2$.

In some circumstances, it might be convenient to express the physical masses $m_{Z_D}^2$ and $m_Z^2$, as well as the angle $\alpha$, as a function of the model mass parameters $m_X^2$ and $\hat m_Z^2$, somehow inverting the equations above. Defining  
$\hat r \equiv m_X^2/\hat m_Z^2$, one finds
\be
m^2_{Z_D}= \frac{1-\epsilon^2 +\hat r- \mathcal{R}}{2(1-\epsilon^2/c_w^2)}\hat m^2_Z~,\qquad
m^2_{Z}= \frac{1-\epsilon^2 + \hat r +\mathcal{R}}{2(1-\epsilon^2/c_w^2)}\hat m^2_Z~,
\label{ZZD}\ee
\be 
\sin 2\alpha = \frac{2t_w\epsilon(1-\epsilon^2/c_w^2)^{1/2}}{\mathcal{R}}~,\qquad
\cos 2\alpha = \frac{1-\epsilon^2/c_w^2-t_w^2\epsilon^2- \hat r}{\mathcal{R}}~,
\ee
where
\be
 \mathcal{R}\equiv 
[(1-\epsilon^2)^2-2(1-\epsilon^2/c_w^2-t_w^2\epsilon^2)\hat r+ \hat r^2]^{1/2}~.
\label{Rratio}\ee

\section{Details of the relic density computation}\label{app:BE}

In this appendix, we provide detailed expressions for the quantities relevant to solve the Boltzmann equations discussed in section~\ref{sec:relic}. We organize the material into three parts: (i) cosmology-related quantities, (ii) computation of the reaction densities, and (iii) explicit formulas for the cross sections and the amplitudes.

\subsection{Cosmology}

The (equilibrium) number densities are defined as
\be \label{eq:neq}
 n_i^{(\rm eq)} =  \frac{g_i}{(2 \pi)^3} \int   f^{(\rm eq)}_i \, d^3p_i  \, , \, \: {\rm with}  \, \: \begin{dcases}
    \, n_i^{\rm eq} =c_i   \frac{\zeta(3) }{\pi^2} \, g_i \, T^3  & \, {\rm for} \quad T \gg m_i \, , \\[10pt]
    \, n_i^{\rm eq} = g_i  \qty(\frac{ m_i \, T }{2 \pi})^{3/2} \exp \qty(-\frac{m_i}{T})  & \, {\rm for} \quad  T \ll m_i \, ,
\end{dcases}
\ee
where $f_i^{(\rm eq)}=(e^{E_i/T} \pm 1)^{-1}$ is the (equilibrium) phase-space distribution function, 
with the plus (minus) sign for fermions (bosons), and we
assumed that all chemical potentials are negligible. 
Here $g_i$ is the number of internal degrees of freedom of species $i$ with mass $m_i$, and the coefficient $c_i$ takes the values $1$ for bosons and $3/4$ for fermions. In the code, when $T \gg m_i$ (relativistic) or $T \ll m_i$ (non-relativistic), we use the analytic expressions for $n_i^{\rm eq}$ given in eq.~\eqref{eq:neq}. However, when $m_i \sim T$, we compute the integral numerically.

The Hubble expansion rate $H$ in a radiation-dominated Universe is given by
\[
H(T) = \sqrt{\frac{8\pi^3 g_\rho(T)}{90}} \, \frac{T^2}{M_{\rm Pl}} \, ,
\]
where $M_{\rm Pl}$ is the Planck mass and $g_\rho(T)$ counts the relativistic degrees of freedom contributing to the energy density at temperature $T$. The comoving number density yield is defined as $Y_i^{\rm (eq)} \equiv n_i^{(\rm eq)}/s$, with the entropy density
\[
s = \frac{2\pi^2}{45} \, g_s(T) \, T^3 \, ,
\]
where $g_s(T)$ counts the relativistic degrees of freedom contributing to the entropy density. Note that $g_s = g_\rho$ unless relativistic species have decoupled from photons, which occurs in the SM only for $T \lesssim m_e$.

\subsection{Reaction densities}

Let us define the reaction densities which enter into the Boltzmann equations.
The general formula for the reaction density $\hat \gamma$ of a process in which a set of particles $\{i\}$ converts into a set $\{j\}$ is given by
\be \label{eq:gamma}
\hat \gamma_{\{i\} \rightarrow \{j\} } \equiv \int   \prod_{i \in \{i\} } d\Pi_i \, f_i  \, \prod_{j \in \{j\} } d\Pi_j \, (2\pi)^4 \delta^4(p_{\{i\}} - p_{\{j\}}) \, |{\cal M}_{\{i\} \rightarrow \{j\}}|^2 \, ,
\ee
where the phase-space element is $d\Pi_i = d^3p_i/[(2\pi)^3 2E_i]$, and the squared invariant amplitude $|{\cal M}_{\{i\} \rightarrow \{j\}}|^2$ is summed over initial and final spins with no average, and divided by $k!$ whenever $k$ identical particles appear in the initial or final state.

Although this expression is completely general, it can be further simplified by noting that the particles belonging to the SM bath follow equilibrium distributions at the common temperature $T$. In the freeze-out case, the DM particles remain in kinetic equilibrium throughout the process, so they also follow their equilibrium distributions. Therefore, one can approximate $f_i \sim f_i^{\rm eq}$. In the freeze-in case, although the DM candidates are not in thermal equilibrium, they are produced from interactions of SM particles in the bath. Consequently, one can still assume that the initial-state particles follow equilibrium distributions.  Finally, one can use the Maxwell-Boltzmann approximation, $f_i^{\rm eq} \simeq e^{-E_i/T}$, as guaranteed e.g.~by kinetic equilibrium, for both the non-relativistic and the relativistic (when $\langle E_i \rangle \sim 3\, T$) regime.

With the replacement $f_i\to e^{-E_i/T}$, the reaction density in eq.~\eqref{eq:gamma} becomes identical to the one for the inverse process, therefore it can be denoted by $\gamma_{\{i\} \leftrightarrow \{j\}}$.
For a generic scattering process  $1+\dots+m\leftrightarrow a+b$, which transforms particle species $\{1,2,\dots,m \}$ into $\{a,b \}$,
the reaction density enters the Boltzmann equation for particle 1
according to
\be
H s x_1 \frac{dY_1}{dx_1} \equiv \dot n_1 + 3H n_1 = \gamma_{1+\dots+m\leftrightarrow a+b}
\left(\frac{Y_a}{Y_a^{eq}}\frac{Y_b}{Y_b^{eq}}-\frac{Y_1}{Y_1^{eq}}\dots \frac{Y_m}{Y_m^{eq}}\right) + {other~processes}~.
\label{masterB}\ee
Such equation describes the evolution of the comoving number density $Y_1\equiv n_1/s$ as a function of $x_1\equiv m_1/T$.

One can derive simplified forms for the reaction densities, depending on the number of particles in the initial and final states. For the case of the \textbf{decay} of a particle 1 into two particles $a$ and $b$, we have
    \be 
    \gamma_{1\leftrightarrow a+b} \equiv 
    n_1^{eq} \, \langle\Gamma_{1\rightarrow a+b}\rangle 
    = n_1^{eq} \, \frac{K_1(x)}{K_2(x)} \, \Gamma_{1\rightarrow a+b}~,
    \ee
where $\Gamma_{1\rightarrow a+b}$ is the decay width, and $K_n$ is the modified Bessel function of second type of order $n$.

For \textbf{2-to-2 processes}, the reaction density simplifies to 
\begin{equation}\label{eq:gam_22}
\gamma_{1+2\leftrightarrow a+b} \equiv 
n_1^{eq}n_2^{eq}  \langle\sigma_{1+2\rightarrow a+b}v\rangle 
= \frac{T}{64\pi^4}\int_{s_{min}}^\infty ds \sqrt{s} K_1\left(\frac{\sqrt{s}}{T}\right)
\hat\sigma_{1+2\leftrightarrow a+b}(s)~,
\end{equation}
where $s_{\rm min} = {\rm max}[(m_1+m_2)^2, (m_a+m_b)^2]$.
The reduced cross-section $\hat \sigma$ is given by
\be \label{eq:redxsec}
\hat \sigma_{1+2 \leftrightarrow a+b}(s) = \frac{g_1 g_2}{c_{12}} \frac{2 \lambda(s, m_1^2, m_2^2)}{s} \sigma_{1+2 \to a+b}(s) \, ,
\ee
in which $\lambda(x,y,z) \equiv (x-y-z)^2 - 4yz$ is the K\"{a}ll\'{e}n-function,  $c_{12}$ is a symmetry factor that equals to $2$ if the initial particles are identical and 1 otherwise, and $\sigma_{1+2 \to a+b}$ is the ordinary $2 \to 2$ cross-section.

Finally, in the case of \textbf{3-to-2 processes}, one can simplify the reaction density under certain assumptions. In the freeze-out scenario, where the relevant temperatures are much smaller than the dark sector masses, the initial particle phase-space is dominated by small momenta, due to the Boltzmann suppression. Consequently, in this non-relativistic regime, the initial particles can be approximated as being at rest, $p_i \simeq (m_i, \vec 0\,)$~\cite{Bernal:2015bla}, and the reaction density can be approximated as
\be \label{eq:gam_32}
\begin{aligned} 
&\gamma_{\scaleto{1+2+3 \leftrightarrow a+b}{6pt}}  \equiv 
n_1^{eq}n_2^{eq}n_3^{eq} \, \ang{\sigma v^2}_{\scaleto{1+2+3 \rightarrow a+b}{6pt}} \\[4pt]
& \simeq  \qty(\, \prod_{\scaleto{i = 1,2,3}{6pt}} \frac{n_i^{\rm eq}}{2 m_i g_i} ) \int d\Pi_a \, d\Pi_b \,
(2 \pi)^4  \,\delta \qty(\sum_{\scaleto{i = 1,2,3}{6pt}} m_i - m_a-m_b) \,  \delta^3( \vec p_a +\vec p_b)    \, |{\cal M}_{\scaleto{1+2+3\rightarrow a+b}{6pt}}|^2_{\rm NR} \, ,
\end{aligned} \\[5pt]
\ee
where the ${\rm NR}$ subscript indicates that the amplitude is computed in the non-relativistic limit, where the initial particle momenta are neglected. Actually, in this approximation the amplitude depends only on the masses and couplings, and it can be factored out of the integral. The remaining phase-space integral can then be solved, to obtain the final expression for the 3-to-2 reaction density. For the cases of interest in this work, the explicit result can be found in eq.~\eqref{eq:sigmav32}.

\subsection{Decay widths, cross sections, amplitudes}

In this subsection, we present the expressions for the decay rates, $2 \to 2$ cross sections, and $3 \to 2$ amplitudes and thermally-averaged cross sections of the processes contributing to the relic density computation.

\paragraph{\underline{Decays}}\mbox{}\\[5pt]
The decay of $Z_D$ into SM fermions, shown in diagram~(b) of figure~\ref{fig:dZDbeq}, is the key process responsible for the depletion of the $Z_D$ population in the early Universe and plays an important role in the dark mediator phenomenology. The general expression for the decay width of a kinetically mixed vector boson, after kinetic mixing and mass diagonalisation in the broken EW phase, is given by~\cite{Curtin:2014cca}
\be \label{eq:ZDdec}
\Gamma_{Z_D \rightarrow \bar f f} = \frac{N_c}{24 \pi m_{Z_D}} \qty[m_{Z_D }^2 \, (g_L^2 + g_R^2)- m_f^2 \, (g_L^2 + g_R^2 - 6 \,  g_L  \, g_R)] \qty(1- \frac{4 m_f^2}{m_{Z_D}^2})^{1/2} \, ,
\ee
where $N_c$ is the number of fermion colours, $m_f$ is the fermion mass, and the couplings $g_{L,R}$ are defined as
\be
g_{L/R} = \, - \frac{g}{c_{w}} \, \qty[ s_\alpha \, (T_{L/R} \, c_{w}^2 - Y_{L/R} \, s_{w}^2) + \frac{\epsilon}{\sqrt{1- \epsilon^2/c_{w}^2}} \, c_{\alpha} \, t_{w} \, Y_{L/R} ],
\ee
with $T_{L/R}$ and $Y_{L/R}$ denoting the weak isospin and hypercharge of the left/right-handed component of the fermion $f$, respectively. Before EW symmetry breaking, the decay width takes the same form as the full expression eq.~\eqref{eq:ZDdec}, with $\alpha = 0$.

In addition to the dark mediator width, the decay width of the SM $Z$ boson, $\Gamma_{Z \rightarrow \bar f f}$, is also relevant for the kinetic mixing diagram discussed in the next subsection. Its expression is the same as in eq.~\eqref{eq:ZDdec}, with the replacements $m_{Z_D} \to m_Z$ and 
\be
g_{L/R}\to g_{L/R}^Z \simeq \, \frac{g}{c_{w}} \, \qty(T_{L/R} \, c_{w}^2 - Y_{L/R} \, s_{w}^2) ,
\ee
where $\mathcal{O}(\epsilon^2)$ corrections have been neglected.

In the limit $m_{Z_D}^2 \ll m_Z^2$, which is the regime mainly considered in this work, the dark mediator becomes predominantly ``photon-like'', and the decay width simplifies to
\be
\Gamma_{Z_D \rightarrow \bar f f} \simeq  N_c \frac{(\epsilon \, q^f_{\rm EM})^2}{12 \pi} \, m_{Z_D} \, \qty(1 +  \frac{2\,m_f^2}{m_{Z_D}^2}) \qty(1- \frac{4\,m_f^2}{m_{Z_D}^2})^{1/2} \, ,
\ee
where $q^f_{\rm EM}$ is the electric charge of $f$.

In the code, we employed these expressions for the fermionic degrees of freedom. However, for $Z_D$ masses below $\sim 2~\mathrm{GeV}$, hadronic decays become relevant due to the onset of the non-perturbative QCD regime. In this region, we accounted for hadronic decays into more than $\sim 20$ channels, which were already implemented in the~\textsc{ReD-DeLiVeR} code~\cite{ReD-DeLiVeR} (for the derivation and further details on the hadronic widths computation, see Ref.~\cite{Foguel:2024lca}).

\paragraph{\underline{$2\to 2$ processes}}\mbox{}\\[5pt] 
The relevant $2\to 2$ processes include the \textbf{kinetic mixing} $s$-channel diagram (a) of figure~\ref{fig:BEQdiagrams}, the \textbf{dark-annihilation} diagrams (b) of figure~\ref{fig:BEQdiagrams}, and the $t$-channel $Z_D$ production from \textbf{top–antitop annihilation}, diagram (a) of figure~\ref{fig:dZDbeq}.

Let us start with the \textbf{kinetic mixing} diagram. Since we consider both freeze-out and freeze-in regimes, and because this process involves SM states, it is important to clarify which degrees of freedom contribute at temperatures above or below the EWSB scale. Consequently, the kinetic-mixing cross section can be defined in two complementary regimes:
\be
\sigma_{XX}^{\, {\rm kinetic}}(s) = \begin{cases}
    \, \sigma_{X^+ X^- \rightarrow \bar f f}^{\, {\rm bEW}}   & \, {\rm for} \quad T > 246 \, {\rm GeV} \, , \\[10pt]
    \,\sigma_{X^+ X^- \rightarrow \bar f f}^{\, {\rm aEW}} & \, {\rm for} \quad T < 246 \, {\rm GeV} \, , \\[10pt]
\end{cases}  
\ee
Before the EWSB scale, the dark mediator mixes with the hypercharge boson, inheriting couplings to the left and right-handed fermion hypercharges. In this regime, diagram (a) of figure~\ref{fig:BEQdiagrams} involves the dark mediator $X^3$ and the hypercharge $B^\mu$ boson as mediators, and the SM states are the left and right-handed fermions (with thermal masses, as described in~\cite{Bringmann:2021sth}). The cross section before EWSB reads
\be
\sigma_{X^+ X^- \rightarrow \bar f f}^{\, {\rm bEW}} (s)= g_X^2 \, \dfrac{  (4 \, m_X^6 - 15 \, m_X^4 \, s + 40 \, m_X^2 \, s^2 + 4 \, s^3) \, \qty({1- \dfrac{4\,m_X^2}{s}})^{1/2} \Gamma_{Z_D \rightarrow \bar f f}(s) }{36\,m_X^4\, \sqrt{s}\, \qty[(s-m_{X}^2)^2 + m_{X}^2 (\Gamma_{Z_D})^2] } \, ,
\ee
where $\Gamma_{Z_D \rightarrow \bar f f}(s)$ denotes the expression in eq.~\eqref{eq:ZDdec}, taken at $\alpha=0$, and evaluated for $m_{Z_D} = \sqrt{s}$, and $\Gamma_{Z_D}$ is the total dark mediator width before EWSB.

For the regime after EWSB, the relevant degrees of freedom are depicted in diagram (a) of figure~\ref{fig:BEQdiagrams}. In the limit where the dark mediator behaves as ``photon-like'' ($m_{Z_D}\ll m_Z$), the cross section can be conveniently written as the sum over three contributions: a $Z$-mediated term, a vector $V = (\gamma, Z_D)$-mediated term (which is a sum over photon and dark-photon exchange), and an interference term,
\begin{equation}
\begin{aligned}
 \sigma_{X^+ X^- \rightarrow \bar f f}^{\, {\rm aEW}}(s) &=  \frac{g_{X}^2
\qty({1- \frac{4\,m_X^2}{s}})^{1/2}}{ 36 \, \sqrt{s} \, c_w^2 \, m_X^4 }
\, \left(\frac{\mathcal{F}_Z(s)}{\scriptstyle{\rm BW}_Z(s)} +  \frac{\mathcal{F}_V(s)}{{\scriptstyle{\rm BW}_V(s) }}  +  \frac{\mathcal{F}_I(s)}{\scriptstyle{{\rm BW}_Z(s)} \, \scriptstyle{{\rm BW}_V(s) }} \right) \, , 
\end{aligned}
\end{equation}
where the Breit-Wigner denominators for the $Z$ and $Z_D$ mediators are defined as
\begin{equation}
\begin{aligned}
\scriptstyle{\rm BW}_Z(s)  & \equiv  (s - m_Z^2)^2 + m_Z^2 \, \Gamma_Z^2 \,, \\
\scriptstyle{{\rm BW}_V(s)}  & \equiv (s - m_{Z_D}^2)^2 + m_{Z_D}^2 (\Gamma_{Z_D}^{\, \rm aEW})^2 \, , 
\end{aligned}
\end{equation}
with $\Gamma_Z \simeq 2.5 $ GeV is the total $Z$-boson width, and $\Gamma_{Z_D}^{\, \rm aEW}$ the total $Z_D$ width after EWSB. 
The $Z$- and $V$-mediated functions, as well as the interference term, are defined as
\begin{equation}
\begin{aligned}
\mathcal{F}_Z(s) & \equiv
  \Gamma_{Z \to \bar f f}(s) \, s \, \Big[ \,( \bm{s_\alpha^2}) \, c_w^2 \, (12  \, m_X^4 + 20 \,  s \, m_X^2 + s^2)  
  \\ & \quad
  + 2 \, ( \bm{s_\alpha \, \epsilon}) \,  c_w s_w \, s \, (10 \, m_X^2 + s)
+ (\bm{\epsilon^2 })\, s_w^2 \, s \, (4 m_X^2 + s)\Big], \\
\mathcal{F}_V(s) & \equiv \bar \Gamma_{Z_D \rightarrow \bar f f}(s) \, \bm{\epsilon^2} \, \big(q^f_{\rm EM} \big)^2 \, c_w^2  \, \qty[ 12  \, m_X^4  \, s
+ 4  \, m_X^2  \, (m_{Z_D}^4 - 7 m_{Z_D}^2 s + 11 s^2)
+ s  \, (m_{Z_D}^2 - 2 s)^2 ], \\
\mathcal{F}_{\rm I}(s) & \equiv
 \bar  \Gamma_{Z_D \rightarrow \bar f f}(s) \qty[ \, q^f_{\rm EM} \, (g_L^Z + g_R^Z) ]  \,\bm{\epsilon} \, \, c_w \, s \,(s - m_{Z_D}^2)(s - m_Z^2) \,  \\
 &\quad \Big[ \bm{\epsilon} \, s_w \qty( m_X^2 (14 s - 4 m_{Z_D}^2) + s \,(2s - m_{Z_D}^2) ) \\
 &\quad + \bm{s_\alpha} \, c_w  \qty( 12\,  m_X^4 - 10 \, m_X^2 (m_{Z_D}^2 - 3s) + s \, (2s - m_{Z_D}^2))  \Big] \, , 
\end{aligned}
\end{equation} \\[0.5pt]
where decay widths as functions of $s$ are evaluated by replacing the mediator mass $m_Z^2$ or $m_{Z_D}^2$ with the energy $s$, and we defined a normalized width $ \bar  \Gamma_{Z_D \rightarrow \bar f f}(s)  \equiv \Gamma_{Z_D \rightarrow \bar f f}(s) /(\epsilon \, q^f_{\rm EM})^2$. Since $s_\alpha \sim \epsilon$, the cross section scales as $\sigma_{X^+ X^- \rightarrow \bar f f} \sim \bm{g_X^2 \, \epsilon^2}$.

Now, let us move on to the \textbf{dark annihilation} case, $X^+X^- \to Z_D Z_D$. The cross section reads
\be
\sigma_{XX}^{\, {\rm dark}}(s) = \dfrac{\bm{g_{X}^4}}{4608 \, \pi \, m_X^4 m_{Z_D}^4 s^2} \, \Big[ \mathcal{A}(s) +\mathcal{B}(s) \Big] \, , 
\ee
where we defined the functions
\begin{equation}
\begin{aligned}
\mathcal{A}(s) &= 
\dfrac{\sqrt{(s - 4 m_X^2 )(s - 4 m_{Z_D}^2)}}{m_X^2 (s - 4 m_{Z_D}^2) + m_{Z_D}^4} 
\Big[ 8 m_X^8 (192 m_{Z_D}^4 + 8 m_{Z_D}^2 s - s^2) \\
&\quad + m_X^6 (4464 m_{Z_D}^6 - 660 m_{Z_D}^4 s + 24 m_{Z_D}^2 s^2 + s^3) \\
&\quad + m_X^4 m_{Z_D}^2 (1622 m_{Z_D}^6 - 660 m_{Z_D}^4 s + 513 m_{Z_D}^2 s^2 - 4  s^3) \\
&\quad - 4 m_X^2 m_{Z_D}^4 (302 m_{Z_D}^6 + 74 m_{Z_D}^4 s - 23 m_{Z_D}^2 s^2 - s^3) \\
&\quad + 4 m_{Z_D}^8 (54 m_{Z_D}^4 - 16 m_{Z_D}^2 s + 3 s^2) \Big],
\\[8pt]
\end{aligned}
\end{equation}
and
\begin{equation}
\begin{aligned}
\mathcal{B}(s) &=\frac{4\, \mathcal{P}(s)}{\,s - 2 m_{Z_D}^2\,} \,
\ln\!\left[
\frac{s - 2 m_{Z_D}^2 + \sqrt{ (s - 4 m_X^2 )(s - 4 m_{Z_D}^2)}}
     {s - 2 m_{Z_D}^2 - \sqrt{ (s - 4 m_X^2 )(s - 4 m_{Z_D}^2)}}
\right] \, ,  \\[8pt] 
\end{aligned}
\end{equation}
with the polynomial $ \mathcal{P}(s)$ given by
\begin{equation}
\begin{aligned}
 \mathcal{P}(s) &= 4 m_X^8 (200 m_{Z_D}^4 - 4 m_{Z_D}^2 s + s^2)
+ 4 m_X^6 m_{Z_D}^2 (464 m_{Z_D}^4 - 139 m_{Z_D}^2 s - 3  s^2) \\
& + m_X^4 m_{Z_D}^4 (661 m_{Z_D}^4 - 668 m_{Z_D}^2 s + 45  s^2) \\
& + 2 m_X^2 m_{Z_D}^4 (158 m_{Z_D}^6 + 74 m_{Z_D}^4 s - 30 m_{Z_D}^2 s^2 - s^3) \\
& + 4 m_{Z_D}^6 (-27 m_{Z_D}^6 + 26 m_{Z_D}^4 s - 6 m_{Z_D}^2 s^2 + s^3) \, .
\end{aligned}
\end{equation}

Finally, the \textbf{top–antitop annihilation} $t$-channel diagram, included in eq.~\eqref{eq:FIbeqZD} for the freeze-in production of the dark mediator $Z_D$, has cross section (before EWSB, since the regime $T>v$ is the relevant one in our UV-dominated scenario)
\begin{equation}
\begin{aligned}
\sigma_{t\bar{t} \to Z_D \Phi}(s)
= & \,
\bm{\epsilon^2} \, \qty(\frac{m_t}{v} \frac{ \, g'}{c_w})^2  \, 
\frac{ \qty[(Y_L^t)^2 +(Y_R^t)^2]}{16 \pi \, s \, (s -m_H^2)^2 \, (s - 4 m_t^2)}
\Bigg[
- 4 \sqrt{s} \, (m_H^2 - 4 m_t^2) (s - m_H^2) \sqrt{s - 4 m_t^2} \\
& + (s- m_H^2 ) \, \qty[m_H^4 - 4 m_H^2 m_t^2 + \qty(s - 4 m_t^2)^2] \, 2 \,
\ln\!\left(
\dfrac{\sqrt{1 -  4 \, m_t^2/s} +1}{\sqrt{1 -  4 \, m_t^2/s} - 1}
\right)
\Bigg] \,,
\end{aligned}
\end{equation}
where $m_t$ and $m_H$ denote the top-quark and Higgs doublet (thermal) masses, respectively, and $Y_{L(R)}^t = 1/6 , (2/3)$ is the left (right)-handed top-quark hypercharge.

Let us conclude by highlighting that, for all the processes discussed above, in the regime where the temperature in the Boltzmann equation exceeds the EWSB scale, SM particle masses are replaced by their thermal masses, as given for instance in Ref.~\cite{Bringmann:2021sth}. This makes the cross sections involving SM initial or final states temperature dependent. However, as emphasised in Ref.~\cite{Becker:2023vwd}, to accurately account for thermal effects requires considering not only thermal masses, but also additional corrections. In the present analysis, we neglect these effects, as they would only induce minor shifts in the freeze-in parameter space, that can be easily compensated e.g.~by a slight adjustment of the reheating temperature, without changing the overall picture depicted in figure~\ref{fig:FImesa}. Nevertheless, we include thermal masses for consistency: without them, spurious diagrams such as the inverse $Z_D$ decay would appear to populate the bath, even though these processes are actually forbidden. In such cases, one should instead consider, for example, the $t$-channel top–antitop annihilation described above for the purpose of $Z_D$ production.

\paragraph{\underline{$3\to 2$ processes}}\mbox{}\\[5pt] 
The relevant $3 \to 2$ or \textbf{SIMP} processes contributing to the Boltzmann equation computation occur for three distinct sets of initial particles: $XXX$, $XXZ_D$, and $Z_D Z_D Z_D$, corresponding to diagrams (c), (d), and (e) of figure~\ref{fig:BEQdiagrams}, respectively. To compute their reaction densities, we employed the non-relativistic approximation given in eq.~\eqref{eq:gam_32}. Considering the generic case where $m_{Z_D} = R \, m_{X}$, we solved the phase-space integral of that equation for each case, obtaining the following results:
\be 
\begin{aligned} \label{eq:sigmav32}
 & \ang{\sigma v^2}_{\scaleto{XXX}{5pt}} & \simeq   & \quad \frac{1}{27} \cdot \dfrac{\sqrt{64 - 20 R^2 + R^4 }}{576 \, \pi}  \, \frac{1}{m_X^3} \,  |{\cal M}_{\scaleto{X^-X^+X^\pm \rightarrow Z_D X^\pm}{6pt}}|^2_{\rm NR} \, \cdot \Theta(2-R) \, , \\[10pt]
 & \ang{\sigma v^2}_{\scaleto{XXZ_D}{6pt}} & \simeq  & \quad \frac{1}{27}  \cdot  \dfrac{\sqrt{4 + 4 R - 3 R^2}}{64 \, \pi \, R \, (R+2)} \,  \frac{1}{m_X^3} \,  |{\cal M}_{\scaleto{X^+X^-Z_D \rightarrow Z_D Z_D}{6pt}}|^2_{\rm NR} \, \cdot \Theta(2-R) \, , \\[10pt]
 & \ang{\sigma v^2}_{\scaleto{Z_D Z_D Z_D}{6pt}} & \simeq  & \quad \frac{1}{27}  \cdot  \frac{\sqrt{9R^2 - 4}}{192 \, \pi \, R^4} \,  \frac{1}{m_X^3} \,  |{\cal M}_{\scaleto{ Z_D Z_D Z_D \rightarrow X^+X^- }{6pt}}|^2_{\rm NR} \, ,
\end{aligned} \\[10pt]
\ee
where $\Theta$ is the Heaviside step function. The non-relativistic amplitudes were computed using \texttt{FeynArts}~\cite{Hahn:2000kx} and \texttt{FeynCalc}~\cite{Shtabovenko:2023idz}, and they are given by
\be 
\begin{aligned}
& |{\cal M}_{\scaleto{XXX}{5pt}}|^2_{\rm NR} & \sim  & \,   \, \dfrac{g_X^6}{m_X^2} \, 
\frac{(R^2 - 16) \, \mathcal{P}_{\scaleto{XXX}{5pt}}(R)}{41472 \, R^2 \, (R^2 - 7)^2 \,  (R^2 - 4)^3  \, (R^2 + 2)^2  \, (R^4 - 20 R^2 - 8)^2} \\[10pt]
& &  &\overset{(R=1)}{=}  \frac{324035}{512} \, \dfrac{g_X^6}{m_X^2} \, , \\[10pt]
& |{\cal M}_{\scaleto{XXZ_D}{6pt}}|^2_{\rm NR} & \sim  & \,  \, \dfrac{g_X^6}{m_X^2} \, 
\frac{(3 R + 2) \, \mathcal{P}_{\scaleto{XXZ_D}{6pt}}(R)}
{1728 \, R^6 \, (R - 2)^3 \, (R + 1)^4 \, (R + 2)^2 \, (2 R^2 - R - 2)^2} \\[10pt]
& & & \overset{(R=1)}{=}  \frac{328285}{256} \, \dfrac{g_X^6}{m_X^2} \, ,  \\[10pt]
& |{\cal M}_{\scaleto{Z_DZ_DZ_D}{6pt}}|^2_{\rm NR} & \sim  & \,  \, \dfrac{g_X^6}{m_X^2} \, 
\frac{(9 R^2 - 4) \, \mathcal{P}_{\scaleto{Z_DZ_DZ_D}{6pt}}(R)}
{16 \, (1 - 3 R^2)^4} \,  \overset{(R=1)}{=}  \frac{27615}{64} \, \dfrac{g_X^6}{m_X^2} \, , \\[10pt]
\end{aligned}
\ee
where the polynomials are defined as
\begin{equation*}
\begin{aligned}
 \mathcal{P}_{\scaleto{XXX}{5pt}}(R) \equiv & \, 49\, R^{30}
- 5036 \, R^{28}
+ 288265 \, R^{26}
- 9\,399\,910 \, R^{24}
+ 173\,332\,892 \, R^{22}
- 1\,871\,704\,096 \, R^{20} \\
&+ 13\,149\,303\,384\,  R^{18}
- 72\,891\,737\,760 \, R^{16}
+ 347\,944\,750\,656 \, R^{14}
- 1\,152\,329\,689\,600 \, R^{12} \\
& + 2\,544\,684\,939\,776 \, R^{10}
- 2\,338\,582\,276\,096 \, R^{8}
+ 3\,415\,661\,940\,736 \, R^{6} \\
& + 6\,330\,651\,934\,720 \, R^{4}
+ 1\,607\,804\,059\,648 \, R^{2}
+ 464\,529\,653\,760 \, , \\[5pt]
\end{aligned}
\end{equation*} 
\begin{equation*} 
\begin{aligned}
  \mathcal{P}_{\scaleto{XXZ_D}{6pt}}(R) \equiv & \, 36 \, R^{20} - 164 \, R^{19} + 2393 \, R^{18} - 376 \, R^{17} + 603476 \, R^{16} + 1319742 \, R^{15} - 1661722 \, R^{14} \\
  & + 1813646 \, R^{13} + 22115229 \, R^{12} - 228802 \, R^{11} - 62811529 \, R^{10} - 44372538 \, R^9 \\
  & + 30747821 \, R^8 + 41485844 \, R^7 + 20010204 \, R^6 + 24517856 \, R^5 + 22218992 \, R^4 \\
  & + 6431040 \, R^3 + 296768 \, R^2 + 788992 \, R + 541696  \, , \\[5pt]
\end{aligned}
\end{equation*} 
\begin{equation*} 
\begin{aligned}
  \mathcal{P}_{\scaleto{Z_DZ_DZ_D}{6pt}}(R) \equiv & \,153\,  R^8 + 306 \, R^7 + 26473 \, R^6 + 1352 \, R^5 - 14080\, R^4 - 1856\,  R^3 + 9744 \, R^2 - 896 \, R + 896  \, . \\[5pt]
\end{aligned}
\end{equation*} 
Note $\ang{\sigma v^2}$'s have dimension minus five, so that the corresponding $\gamma$'s defined by eq.~\eqref{eq:gam_32} have dimension four, as required. The $XXX$ channel was already consider in Ref.~\cite{Choi:2019zeb} for the case $R=1$, and our general result agrees well with their value.

\bibliographystyle{utphys}
\bibliography{LightDarkSector.bib}

\end{document}